\documentclass[12pt]{article}
\usepackage{graphicx}
\usepackage{rotating}
\usepackage{url}
\usepackage{amsmath}
\usepackage{hhline}
\usepackage{amssymb}
\usepackage{times}
\usepackage{cite}
\usepackage{array}
\usepackage{multirow}
\usepackage{ifpdf}

\usepackage{color}
\definecolor{rltred}{rgb}{0.75,0,0}
\definecolor{rltgreen}{rgb}{0,0.5,0}
\definecolor{rltblue}{rgb}{0,0,0.75}

\pdfoutput=1       
\pdftrue

\ifpdf
\usepackage{thumbpdf}
\usepackage{epstopdf}

\fi

\newlength{\dinwidth}
\newlength{\dinmargin}
\begin{document}
\begin{titlepage}

\noindent
Date:        \today       \\
                
\vspace{2cm}

\begin{center}
\begin{Large}

{\bf $p$-values for Model Evaluation}

\vspace{2cm}

F. Beaujean$^1$, A. Caldwell$^1$, D. Koll\'ar$^2$, K. Kr\"oninger$^3$ \\

\end{Large}

\vspace{2cm}
$^1$Max-Planck-Institut f\"ur Physik, M\"unchen, Germany \\
$^2$CERN, Geneva, Switzerland \\
$^3$Georg-August-Universit\"at, G\"ottingen, Germany

\end{center}

\vspace{2cm}

\begin{abstract}
Deciding whether a model provides a good description of data is often based on a goodness-of-fit criterion summarized by a $p$-value.  Although there is considerable confusion concerning the meaning of $p$-values, leading to their misuse, they are nevertheless of practical importance in common data analysis tasks.  We motivate their application using a Bayesian argumentation.  We then describe commonly and less commonly known discrepancy variables and how they are used to define $p$-values.  The distribution of these are then extracted for examples modeled on typical data analysis tasks, and comments on their usefulness for determining goodness-of-fit are given.
\end{abstract}
\end{titlepage}

\section{Introduction}
 Progress in science is the result of an interplay between model building and the testing of models with experimental data.  In this paper, we discuss model evaluation and focus primarily on situations where a statement is desired on the validity of a model without explicit reference to other models. We introduce different discrepancy variables\footnote{A discrepancy variable~\cite{ref:discrepancy} is an extension of classical test statistics to allow possible dependence on unknown (nuisance) parameters.} for this purpose and define $p$-values based on these. We then study the usefulness of the $p$-values for passing judgments on models with a few simple examples reflecting commonly encountered analysis tasks. 
 
In the ideal case, it is possible to calculate the degree-of-belief in a model based on the data. This option is only available when a complete set of models and their prior probabilities can be defined. However, the conditions necessary for this ideal case are usually not met in practice.  We nevertheless often want to make some statement concerning the validity of the model(s).  We then are left with using probabilities\footnote{We use {\it probability} to also include {\it probability density}.} of data outcomes assuming the model to try to make some judgments.   These probabilities can be determined deductively since the model is assumed, and therefore frequencies of possible outcomes can be produced within the context of the model. These can then be used to produce frequency distributions of discrepancy variables, and $p$-values (one-sided probabilities for the discrepancy variables) can be calculated using the distributions and the observed values. The use of $p$-values has been widely discussed in the literature~\cite{ref:pvalues} and many authors have commented that $p$-values are frequently misused in claiming support for models~\cite{ref:Schervish}. We give a Bayesian argumentation for the use of $p$-values to make judgments on model validity, and it is in this Bayesian sense that we will use $p$-values.
 
We start with a review of model testing in a Bayesian approach when an exhaustive set of models is available and a coherent probability analysis is possible. We then move to situations where this is not the case, and review some possible choices of discrepancy variables for Goodness-of-Fit (GoF) tests. Next, we define the $p$-values for the discrepancy variables and evaluate their usefulness for several example data sets.   Note that we omit a discussion of Bayes' factors since these require the definition of at least two models for evaluation.

\section{Full set of models available}
\subsection{Formulation}
Assume we have a complete set of models available for describing the data, such that we are sure the data can be 
described by one of the models available. The models can be used to calculate `direct probabilities'; i.e., relative 
frequencies of possible outcomes of the results were one to reproduce the experiment many times under identical 
conditions.    The probability of a model, $M$, is denoted by $P(M)$, with
\begin{equation}
\label{eq:norm}
0 \leq  P(M)  \leq 1 \; ,
\end{equation}
while the probability densities of the model parameters, $\vec{\theta}$, are typically continuous functions\footnote{Note that there is in principle no mathematical distinction between {\it model} and  {\it parameters}.  In practice, we distinguish them because models are fixed constructs for which we evaluate the degree-of-belief that the model is correct, whereas parameters can take on a range of values and the analysis is used to extract a degree-of-belief for a particular value.}.  In the 
Bayesian approach, the quantities $P(M)$ and $P(\vec{\theta}|M)$ are treated as probabilities, although they are not frequency distributions and are more accurately described as `degrees-of-belief' 
(DoB). This DoB is updated by comparing data with the predictions of the models. 
$P(M=A)=1$ represents complete certainty that $A$ is the model which describes the data, and $P(A)=0$ represents 
completely certainty that $A$ is not the correct model.

The procedure for updating our DoB using experimental data is
\begin{equation}
\label{eq:Bayes}
P_{i+1}(\vec{\theta},M|\vec{D}) \propto P(\vec{x}=\vec{D}|\vec{\theta},M) P_{i}(\vec{\theta},M) \; ,
\end{equation}
where the index on $P$ represents a `state-of-knowledge'.    The posterior probability density function, $P_{i+1}$, is 
usually written simply as $P$, and the prior is written as $P_0$. The posterior describes the state of knowledge {\it after} the experiment is analyzed.  The quantity  $P(\vec{x}=\vec{D}|\vec{\theta},M)$ represents a probability of getting the data $\vec{D}$ given the model and parameter values, and can usually be defined in a number of ways (see for example Section~\ref{sec:expo}).

Normalizing \ref{eq:Bayes}, and using
$$
 P(\vec{D}) = \sum_M \int P(\vec{x}=\vec{D}|\vec{\theta},M) P_{0}(\vec{\theta},M) d\vec{\theta}
$$
yields
\begin{equation}
P(\vec{\theta},M|\vec{D}) =\frac{P(\vec{D}|\vec{\theta},M) P_0(\vec{\theta},M)}
{P(\vec{D})} \; .
\end{equation}
This is the classic equation due to Bayes and Laplace~\cite{ref:Bayessummary}.

Models can be compared and the DoB in a model can be obtained using
\begin{eqnarray}
\label{eq:probab}
P(M|\vec{D})&=&\int P(\vec{\theta},M|\vec{D}) d\vec{\theta} \;\; .
\end{eqnarray}
This evaluation requires the specification of a full set of models and a definition of the prior beliefs such that
\begin{equation}
\sum_M P_0(M)=1 \;\; .
\end{equation}
It is very sensitive to the definitions of the priors when the data are not very selective.  

\subsection{Example with full set of models}
An example where this approach was used is given in~\cite{ref:ACKK}, and is reviewed here.  The analysis is to be performed on an observed energy spectrum, for which we can form a background-only hypothesis, or a background+signal hypothesis. An example is the search for neutrinoless double beta decay. 

In the following, $H$ denotes the hypothesis that the observed
spectrum is due to background only; the negation, interpreted here as
the hypothesis that the signal process contributes to the
spectrum\footnote{Since the shape of the background spectrum is
assumed to be known the case of unknown background sources
contributing to the measured spectrum is ignored. However, the overall
level of background is allowed to vary.}, is labeled
$\overline{H}$. The posterior probabilities for $H$ and $\overline{H}$ can be
calculated using 

\begin{equation}
P(H|\textrm{spectrum}) = \frac{P(\textrm{spectrum}|H) \cdot P_{0}(H)}{P(\mathrm{spectrum})} 
\label{eqn:pH}
\end{equation} 

and 

\begin{equation}
P(\overline{H}|\textrm{spectrum}) = \frac{P(\textrm{spectrum}|\overline{H}) \cdot P_{0}(\overline{H})}
{P(\mathrm{spectrum})} \; .
\label{eqn:pHbar}
\end{equation} 

\noindent The probability $P(\mathrm{spectrum})$ is

\begin{equation}
P(\textrm{spectrum}) = P(\textrm{spectrum}|H) \cdot P_{0}(H) + P(\textrm{spectrum}|\overline{H}) \cdot P_{0}
(\overline{H}) \;\; .
\end{equation} 

The probabilities for a given spectrum can be calculated based on assumptions on the signal strength and background shape as described in~\cite{ref:ACKK}.  Evidence for a signal or a discovery is then decided based on the resulting value for $P(H|\textrm{spectrum})$. 

In this analysis, it was assumed that the observed spectrum must come from either the background model or a combination of the background and double beta decay signal.  The probability of each case is evaluated and conclusions are drawn from these probabilities.

\section{Incomplete set of models}

In most cases, we analyze data without having an exhaustive set of models available, but nevertheless want to reach 
conclusions on how well the models account for the data.  This information can be used, for example, to guide the search for new models.  In the example given above, it is possible that there are unknown sources of background for which predictions are not possible before the experiment is performed.  The quantities $ P(\textrm{spectrum}|H)$ and $ P(\textrm{spectrum}|\overline{H})$ could be individually examined and, if both are on the small side of the expected distribution, doubts concerning the completeness of our set of models could arise.  

\subsection{General approach to GoF tests}
For a given model, we can define one or more discrepancy variable(s) and calculate the expected frequency distribution of this discrepancy variable.  If the discrepancy variable is well chosen, then the distribution for a `good' model should look significantly different than for a `bad' model.  Finding the discrepancy variable in the region populated by incorrect models then gives us cause to think our model is not adequate.
\subsection{Definition of a $p$-value}
A $p$-value is the probability that, in a future experiment, the discrepancy variable will have a larger value (indicating greater deviation of the data from the model) than the value observed, assuming that the model is correct and all experimental effects are perfectly known.  In other words, not only is the model the correct one to describe the physical situation, but correct distribution functions are used to represent data fluctuations away from the `true values'.  We will focus on GoF tests for the underlying model, but it should be clear that incorrect formulations of the data fluctuations will bias the $p$-value distributions to lower (if the data fluctuations are underestimated) or higher (if the data fluctuations are overestimated) values.

In general, any discrepancy variable which can be calculated for the
observations can be used to define a $p$-value. We use
$R(\vec{x}|\vec{\theta},M)$ and $R(\vec{D}|\vec{\theta},M)$ to denote
discrepancy variables evaluated with a possible set of observations
$\vec{x}$ for given model and parameter values, and for the observed
data, $\vec{x}=\vec{D}$, respectively. To simplify the notation, we
will occasionally drop the arguments on $R$ and use $R^D$ to denote
the value of the discrepancy variable found from the data set at
hand. $R$ can be interpreted as a random variable (e.g., possible
$\chi^{2}$ values for a given model), whereas $R^D$ has a fixed value
(e.g., the observed $\chi^{2}$ derived from the data set at hand).

  Assuming that smaller values of $R$ imply better agreement between the data and model predictions, the definition of $p$ (for continuous distributions of $R$) is written as:
\begin{equation}
p=\int_{R>R^D}  P(R|\vec{\theta},M)  dR \;\; .
\end{equation}

The quantity $p$ is the `tail-area' probability to have found a result with $R(\vec{x})>R^D$, assuming that the model $M$ and the parameters $\vec{\theta}$ are valid.   If the modeling is correct (including that of the data fluctuations), $p$ will have a flat probability distribution between $[0,1]$. For discrete distributions of $R$, the integral is replaced by a sum, the $p$-value distribution is no longer continuous, and the cumulative distribution for $p$ will be step-like.

If the existing data are used to modify the parameter values, the
extracted $p$-value will be biased to higher values.  The amount of
bias will depend on many aspects, including the number of data points,
the number of parameters, and the priors.  We can remove the bias for the
number of fitted parameters in $\chi^2$ fits by evaluating the
probability of $R=\chi^2$ for $N-n$ degrees-of-freedom,
$P(\chi^2|N-n)$, where $N$ is the number of data points and $n$ is the
number of parameters fitted~\cite{ref:Eadie}, when
\begin{itemize}
\item the data fluctuations are Gaussian and independent of the parameters, 
\item the function to be compared to the data depends linearly on the
  parameters, and
\item the parameters are chosen such that $\chi^{2}$ is at its global
  minimum.
\end{itemize}

\noindent
In general, the bias introduced by the number of fitted parameters becomes small if $N\gg n$.

  $p$-values cannot be turned into probabilistic statements about the
model being correct without priors, and statements of `support' for a
model directly from the $p$-value behave
`incoherently'~\cite{ref:Schervish}.  Furthermore, approximations used
for the distributions of the discrepancy variables, biases introduced
when model parameters are fitted and difficulties in extracting
reliable information from numerical algorithms used to evaluate the
discrepancy variable (see section~\ref{sec:comparisonpvalues}) further
complicate their use.  $p$-values should therefore be handled with
care. Nevertheless, we discuss the use of $p$-values to make
judgments about the models at hand.  The judgment will be based on a
sequence of considerations of the type:
\begin{itemize}
\item the $p$-value distribution for a good model is expected to be (reasonably) flat between $[0,1]$;
\item the $p$-values for bad models usually have sharply falling distributions starting at $p=0$;
\item small $p$-values are worrisome; if we know that other models can be reasonably constructed which would have higher $p$-values, then a small $p$-value for the model under consideration indicates that we may have picked a poor model;
\item if the $p$-value is not too small, then our model is adequate to describe the existing data.
\end{itemize}

\subsection{Bayesian argumentation}
\label{sec:pBayes}
We contend that the use of $p$-values for evaluation of models as just described is essentially Bayesian in character.  Following the arguments given above, assume that the $p$-value probability density for a good model, $M_0$, is flat, 
$$P(p|M_0)=1 \;\; ,$$
and that for poor models, $M_i \;\; (i=1\dots k)$, can be represented by
$$P(p|M_i) \approx c_i e^{-c_i p}$$
where $c_i \gg 1$ so that the distribution is strongly peaked at $0$ and approximately normalized to $1$.
The DoB assigned to model $M_0$ after finding a $p$-value $p$ is then
\begin{equation}
P(M_0|p) = \frac{P(p|M_0)P_0(M_0)}{\sum_{i=0}^k P(p|M_i)P_0(M_i)} \;\; .
\end{equation}
If we take all models to have similar prior DoBs, then
$$P(M_0|p) \approx \frac{P(p|M_0)}{\sum_{i=0}^k P(p|M_i)} \;\; .$$
In the limit $p\rightarrow 0$, we have
$$P(M_0|p) \approx \frac{1}{1+\sum_{i=1}^{k} c_i} \ll 1 $$
while for $c_i p\gg 1 \;\; \forall  i>0$
$$P(M_0|p) \approx 1\;\; .$$

Although this formulation in principle allows for a ranking of models, the vague nature of this procedure indicates that any model which can be constructed to yield a reasonable $p$-value should be retained. A further consideration is that the correct distributions for the data fluctuations are often not known (due to the vague nature of systematic uncertainties) and best guesses are used.  This will generally also lead to non-flat $p$-value distributions for good models.  

Scientific prejudices (Occam's razor, elegance or esthetics, etc.) will influence the decision and act as a guide in selecting the `best' model in cases where several good models are available.

\subsection{Discrepancy variables considered in this paper}

\subsubsection{$\chi^2$ test for data with Gaussian uncertainties}
\label{sec:chi2gauss}

For uncorrelated data assumed to follow Gaussian probability distributions relative to the model predictions, the $\chi^2$ value is a natural discrepancy variable for a GoF test:
\begin{equation}
R_{G} = \chi^2 = \sum_i^N \frac{\left(y_i - f(x_i|\vec{\theta},M)\right)^2}{\sigma_i^2}
\end{equation}
where the $N$ data points are given by $\{\left(x_i,y_i\right)\}$, and the prediction of the model for $y_i$ is $f(x_i|\vec{\theta},M)$. The modeling of the data fluctuations uses fixed standard deviations $\sigma_i$. 

If the parameters of the model are fitted to the data by minimizing $\chi^{2}$ and there are $n$ parameters we replace $\vec{\theta}$ in the formula above with $\vec{\theta}^*$. The $\chi^2$ probability distribution is evaluated for $N-n$ `degrees-of-freedom'.   In the special case where $f$ is linear in the parameters, this procedure again yields a flat $p$-value distribution between $[0,1]$.

\subsubsection{Runs test}

The standard $\chi^2$ test does not take into account clustering of data below or above expectations.
To detect clusters the ordered set of $N$ observations $\{\left(x_i,y_i\right)\}$ is partitioned into subsets containing the success and failure runs (defined as sequences of consecutive $y_i$ above or below the expectation from the model, $f(x_i|\vec{\theta},M)$, respectively). Several discrepancy variables based on success runs can be found in the literature~\cite{ref:runs_classic} but these do not take into account the size of the deviation, $y_i - f(x_i|\vec{\theta},M)$. Recently, a discrepancy variable for runs incorporating this extra information was proposed for ordered data with Gaussian fluctuations~\cite{ref:runs}.  

Let $A_j$ denote the subset of the observations of the $j^{th}$
success run. The weight of the $j^{th}$ success run is then taken to
be
$$\chi^2_{\mathrm{run}, j} =\sum_{i=j_1}^{j_1+N_j-1} \frac{\left(y_i - f(x_i|\vec{\theta},M)\right)^2}{\sigma_i^2}$$
where the sum over $i$ covers the $\{\left(x_i,y_i\right)\} \in A_j$ and $N_j$ is the length of the run. The discrepancy variable
is the largest weight of any run
$$R_{sr} = \max_{j} \chi^2_{\mathrm{run}, j} \;\; .$$

The explicit distribution of $R_{sr}$, used to define the $p$-value,
$p=P(R_{sr}>R^D_{sr}|N)$, is given in~\cite{ref:runs} for the case
when $(\vec{\theta},M)$ are fully specified (no fitting). A similar
discrepancy variable can be defined for failure measurements,
$R_{fr}$. 

To illustrate the definition we present a simple example. Suppose
$N=5$ observations at $x$ positions $\left(1,2,3,4,5\right)$ with
standardized residuals $(y_i - f(x_i|\vec{\theta},M))/\sigma_i$ given
by \linebreak $\left( 0.3, -0.1, -0.8, 0.4, 0.2 \right).$ Then there
are two success runs $A_1 = \{ (1,0.3) \}$, \linebreak $A_2 = \{ (4,0.4),
(5,0.2)\}$ and we find $R_{sr} = 0.16+0.04= 0.2$ due to the second
run. Similarly, for the single failure run, $R_{fr}= 0.65$.

\subsubsection{$\chi^2$ tests for Poisson distributed data}
\label{sec:chi2poisson}

In the case of binned, Poisson distributed data, the quantity
\begin{equation}
R_{P}=\chi^2_P=\sum_{i=1}^{N_{b}}\frac{\left(m_{i}-\lambda_{i}\right)^{2}}{\lambda_{i}}\label{eq:Poisson Pearson}
\end{equation} 
can be used as a discrepancy variable where $N_{b}$ is the number of
bins, the $m_i$ are the numbers of measured events and
$\lambda_i=\lambda_i(\vec{\theta},M)$ are the model expectations for
event bin $i$; for an expectation $\lambda_i$ the expected variance is
$\lambda_i$.  This Pearson $\chi^2$ statistic~\cite{ref:Pearson} was
originally proposed for multinomial data but has found wide use in
analyzing Poisson distributed data. Rather than using the probability
distribution of this discrepancy variable directly, $P(\chi^2|N_{b})$
is often used as an approximation for $P(R_{P})$\footnote{The use of
  this approximation probably dates back to a time when complicated
  numerical calculations were not possible.}. The distribution of
$R_{P}$ has been shown to asymptotically reach a $\chi^2$ distribution
for multinomially distributed data, giving some justification for this
procedure. Practically, this is the case for data with a large number
of entries $m_{i}$ in all bins. When parameters are first estimated
from a fit to the data, the parameter values which minimize $R_{P}$
are used to calculate the $\lambda_i$, and the $p$-value is evaluated
using $P(\chi^2|N_{b}-n)$.
 
It is often seen that the \emph{expected} weight, $\lambda_{i}$, is
replaced with the \emph{observed }weight, $m_{i}$, in the denominator
(Neyman $\chi^2$~\cite{ref:Neyman}). The discrepancy variable is then
\begin{equation}
R_{N} = \chi^2_N=\sum_{i=1}^{N_{b}}\frac{\left(m_{i}-\lambda_{i}\right)^{2}}{m_{i}} \;\; . \label{eq:Poisson obs}\end{equation}
Again, rather than using the probability distribution of this
discrepancy variable directly, it is assumed that $R_{N}$ has a
distribution which approximates a $\chi^{2}$ distribution with $N_{b}$
degrees of freedom and $P(\chi^2|N_{b})$ is used as an approximation
for $P(R_{N})$.  In cases where $m_{i}=0$, practitioners of this
approach set $m_i=1$ to avoid divergence.  Sometimes bins with $m_i=0$
are ignored, which can lead to very misleading results since finding
$m_i=0$ is valuable information.  When parameters are first estimated
from a fit to the data, the parameter values which minimize $R_{N}$
are used, and the $p$-value is evaluated using $P(\chi^2|N_{b}-n)$.

Note that in both of these cases, we do not expect flat $p$-value
distributions since only approximations are used for $P(R_{P/N})$.
The deviations from flatness are expected to be greatest when small
event numbers are present in the data sets.

\subsubsection{Likelihood ratio test for Poisson distributed data}
\label{sec:likeratio}

Another option for a Poisson model is based on the (log of) the
likelihood ratio~\cite{ref:PDG} (sometimes referred to as the Cash
statistic~\cite{Cash:1979vz})
\begin{equation}
R_{C}=2 \log\frac{P(\vec{x}|\lambda_i=m_i)}{P(\vec{x}|\lambda_i=\lambda_i(\vec{\theta}))}= 2\sum_{i=1}^{N_{b}}\left[\lambda_{i}-m_{i}+m_{i}\log\frac{m_{i}}{\lambda_{i}}\right] \;\;.
\label{eq: Poisson likelihood ratio}
\end{equation}
In bins where $m_i=0$, the last term is set to $0$.  Again, rather
than using the probability distribution of this discrepancy variable
directly, since $R_{C}$ has a distribution which approximates a
$\chi^{2}$ distribution with $N_{b}$ degrees of freedom for large
$\lambda_i$, $P(\chi^2|N_{b})$ is used as an approximation for
$P(R_{C})$. The validity of this approximation is critically discussed
in~\cite{Protassov:2002sz}. When parameters are first estimated from a
fit to the data, the parameter values which maximize the likelihood
are used, and the $p$-value is evaluated using
$P(\chi^2|N_{b}-n)$. Note that the distribution of $R_{C}$
asymptotically converges faster to the $\chi^2$ distribution than the
distribution of $R_{P}$ (see \cite{ref:BC}).

\subsubsection{Probability of the data test}
\label{sec:probpoiss}

Any probability of the data can be chosen as the discrepancy variable:
$$R_{L}=P(\vec{x}|\vec{\theta},M) \;\; .$$
In this case, larger values of $R_{L}$ imply better agreement with the data.
The probability $P(\vec{x}|\vec{\theta},M)$ can be used to extract the probability for $R_{L}$ as
$$P(R_{L})dR_{L} = \int_{R_{L}<P(\vec{x}|\vec{\theta},M)<R_{L}+dR_{L}}P(\vec{x}|\vec{\theta},M) d\vec{x} \;\; .$$
 An example of how this is done numerically for Poisson distributed data and using 
$$P(\vec{m}|\vec{\theta},M)=\prod_i^{N_{b}} \frac{e^{-\lambda_i}\lambda_i^{m_i}}{m_i!}$$
is given in the Appendix.  Once we have $P(R_{L}|\vec{\theta},M)$ we can then evaluate 
$$p=\int_{R_{L}<R^D_{L}}  P(R_{L})  dR_{L} \;\; $$
where $R^D_{L}$ is the value observed with the data set at hand.

In a model with Gaussian uncertainties where we use a product of Gaussian densities,
$$P(\vec{x}|\vec{\theta},M)\propto \prod_i^N P(x_i|\mu_i,\sigma_i)$$
where $x_i \sim \mathcal{N}(\mu_i,\sigma_i)$,
then taking $R_{L}=P(\vec{x}|\vec{\theta},M)$ is equivalent to the usual $\chi^2$ test. 

If the model parameters are first fitted using the data, we propose the following correction for the number of fitted parameters:
\begin{itemize}
\item calculate the $p$-value taking $R^D_{L}=P(\vec{x}=\vec{D}|\vec{\theta}^*,M)$, but assuming a simple hypothesis (no fitted parameters);
\item calculate the $\chi^2$ value which corresponds to this $p$-value using the inverse $\chi^{2}$ distribution corresponding to $N$ degrees of freedom;
\item recalculate the $p$-value using the $\chi^2$ value and $N-n$ degrees of freedom.
\end{itemize}
This procedure is valid for the case where we have Gaussian uncertainties, but is ad-hoc for other cases.  We test its usefulness below.

Care must be taken in using $R_{L}$ as a discrepancy variable, particularly when it is written as the product of individual probability densities. The overall shape of the distribution is potentially not tested, and large $p$-values can be produced with incorrect model choices.  An example of this is given below in Section \ref{sec:expo}.

\subsubsection{Partial/Prior/Posterior-predictive $p$-value}

Rather than using the parameters at the mode of the posterior, it is also possible to define a $p$-value by averaging over the parameter values according to a probability distribution~\cite{ref:partial_posterior}. 

For the posterior-predictive case~\cite{ref:posterior}, assuming we are using a probability of the
data as discrepancy variable,
\begin{equation}
\label{eq:posterior_predictive}
p=\int \left[ \int_{R_{L}<R_{L}^D}  P(R_{L}|\vec{\theta},M) dR_{L} \right] P(\vec{\theta}|\vec{D},M)  d\vec{\theta} \;\; .
\end{equation}

While the partial posterior-predictive $p$-value~\cite{ref:partial_posterior} has the desirable property of  a flat distribution on $[0,1]$, at least in the large sample limit as $N\to\infty$, it is not known in general how to compute it for realistic problems. Furthermore, the numerical effort required to evaluate the double integral in \eqref{eq:posterior_predictive} quickly becomes prohibitive. Therefore, these $p$-value definitions are not considered here despite their appeal from a Bayesian perspective.

\subsubsection{Johnson test}
\label{sec:chi2johnson}

Johnson~\cite{ref:Johnson} proposed a modification of the $\chi^2$ definition to take into account the posterior probability density for the parameters of a model.  Rather than evaluating the probability of the data at fixed values of the parameters, the parameters are given values according to their probability density after evaluating the data. Johnson's statistic, $R_{J}$, is expected to behave asymptotically as a $\chi^2$ distribution with $N_{b}-1$ degrees of freedom, where $N_{b}$ is a number of bins to be defined, regardless of the number of parameters. Imagine the data is given by a vector of values $\vec{x}$ and these values are expected to deviate from the model prediction according to individual probabilities $P(x_j|\vec{\theta},M)$.  The Johnson prescription is to define $N_{b}$ bins, where bin $i$ contains a probability $P_i$. Intervals $\Delta x_{i,j}$ are defined for each data point $j$ via
$$P_i(\vec{\theta})=\int_{\Delta x_{i,j}}P(x_{j}|\vec{\theta},M) dx_j  \;\; .$$
The intervals $\Delta x_{i,j}$ cover the full range of possible values for each $x_j$ and are ordered so that they include increasing values of $x$.   The definition of the intervals depends on the values of the parameters $\vec{\theta}$ and vary for each data point $j$.  The parameter values are to be sampled from the full posterior probability $P(\vec{\theta}|\vec{D},M)$.  For each  $\vec{\theta}$, a discrepancy variable $R_{J}$ is calculated according to
\begin{equation}
R_{J}=\sum_{i=1}^{N_b}\frac{\left(m_{i}-NP_{i}(\vec{\theta})\right)^{2}}{NP_{i}(\vec{\theta})} 
\label{eq: Johnson R^B}
\end{equation}
where $m_i$ is the actual number of data points falling within the intervals $\Delta x_{i,j}$ and $N$ is the total number of data points in the data set. 
In \cite{ref:Johnson} it is shown that asymptotically $R_{J}$ is  distributed as a $\chi^2$ variable with $N_{b}-1$ degrees of freedom, regardless of the number of fitted parameters $n$. Hence the $p$-value is calculated using $P(\chi^2|N_{b}-1)$.

For data where the $\vec{x}$ follow continuous probability densities the bins are usually chosen to have equal probabilities.  The number of bins to be chosen is given by a rule of thumb due to Mann/Wald \cite{ref:MannWald} with a modification for small number of bins such that there are at least three:
\[N_{b}=\left\lfloor e^{0.4\cdot\log\left(N\right)}+2\right\rfloor .\]
This means by default we have $N_{b}\left(N=25\right)=5$, $N_{b}\left(N=100\right)=8$, $N_{b}\left(N=1,000\right)=17$. 

In case the values of $\vec{x}$ follow a discrete distribution it is usually not possible to have equal probabilities for all $P_i$.  In this case, a randomization procedure is used to allocate data points to bins.

\section{Test of $p$-value definitions}
In the following, we test the usefulness of the different $p$-values given above by looking at their distributions for specific examples motivated from common situations faced in experimental physics.  We first consider a data set which consists of a background known to be smoothly rising and, in addition to the background, a possible signal.  This could correspond for example to an enhancement in a mass spectrum from the presence of a new resonance.  The width of the resonance is not known, so that a wide range of widths must be allowed for.  Also, the shape of the background is not well known.  We do not have an exhaustive set of models to compare and want to look at GoF's for models individually to make decisions. In this example, we will first assume that distributions of the data relative to expectations are modeled with Gaussian distributions.  We then consider the same problem with small event numbers, so that Poisson statistics are appropriate, and again test our different $p$-value definitions.    These examples were also discussed in ~\cite{ref:BAT}.  Finally, we consider the case of testing an exponential decay law on a sample of measured decay times.

\clearpage
\pagebreak

\subsection{Data with Gaussian uncertainties}
\subsubsection{Definition of the data}

\begin{figure}[h!tbp] 
   \centering
   \includegraphics[width=3.5in]{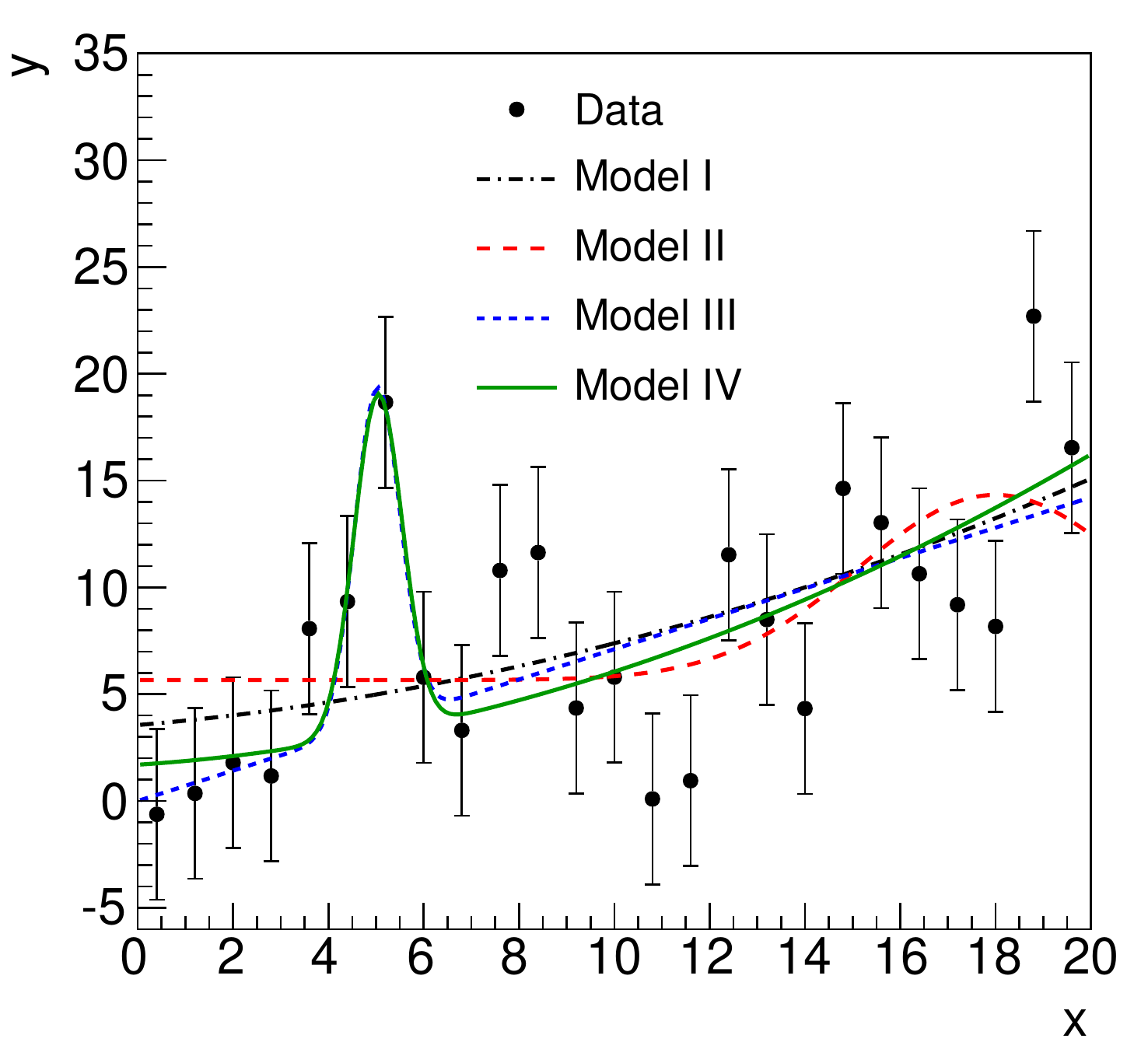}
   \caption{Example data set for the case $N=25$ with Gaussian fluctuations.  The fits of the four models are superposed on the data.}
   \label{fig:Sample-data-set}
\end{figure}

A typical data set is shown in Fig.~\ref{fig:Sample-data-set}. It is generated from the function\begin{equation}
f(x_{i})=A+B\, x_{i}+C\, x_{i}^{2}+\frac{D}{\sigma\sqrt{2\pi}}e^{-\frac{(x_{i}-\mu)^{2}}{2\sigma^{2}}}\,,\label{eq:true 
function}\end{equation}
with parameter values ($A=0$, $B=0.5$, $C=0.02$, $D=15$, $\sigma=0.5$, $\mu=5.0$). The $y_i$ are generated from $f(x_i)$ as
$$y_i=f(x_i)+z_i \cdot 4$$
where $z_i$ is sampled according to $\mathcal{N}(0,1)$.

The $x$ domain is $\left[0,\,20\right]$ and two cases were considered: $N=25$ data points evenly sampled in the range, and $N=100$ data points evenly sampled in the range. The experimental resolution (Gaussian with width $4$) is assumed known and correct. Ensembles consisting of 10,000 data sets with $N=25$ or $N=100$ data points each were generated and four different models were fitted to the data. Table~\ref{tab:models} summarizes the parameters available in each model and the range over which they are allowed to vary. In all models, flat priors were assumed for all parameters for ease of comparison between results. 
\begin{table}[htdp]
\begin{center}
\begin{tabular}{|cl|c|cc|cc|}
\hline
&            &        &\multicolumn{2}{c|}{Small range} &\multicolumn{2}{c|}{Large range} \\
& Model & Par & Min & Max  & Min & Max  \\
\hline
I.  & $A_{\rm I} + B_{\rm I} \, x_{i} + C_{\rm I} \, x_{i}^{2}$  & $A_{\rm I}$ & $\phantom{-}0$ & $\phantom{00}5$ & -50 & 200\\
    & & $B_{\rm I}$ & \phantom{-}0 & \phantom{00}1.2 &-50 & 200 \\ 
    & & $C_{\rm I}$ & -0.1           & \phantom{00}0.1&-50 & 200 \\ 
\hline
II. & $A_{\rm II} + \frac{D_{\rm II}}{\sigma_{\rm II} \, \sqrt{2 \pi}} e^{- \frac{(x_{i}-\mu_{\rm II})^{2}}{2\sigma_{\rm II}^{2}}}$ & $A_{\rm II}$ & \phantom{-}0 & \phantom{0}10 &-50 & 200 \\ 
 & & $B_{\rm II}$      & \phantom{-}0 & 200            &-50 & 200 \\
 & & $\mu_{\rm II}$    & \phantom{-}2 & \phantom{0}18  &0 & 50 \\ 
 & & $\sigma_{\rm II}$ & \phantom{-}0.2 & \phantom{00}4  &0 & 20 \\
\hline
III. & $A_{\rm III} + B_{\rm III} \, x_{i} + \frac{D_{\rm III}}{\sigma_{\rm III} \, \sqrt{2 \pi}} e^{- \frac{(x_{i}-\mu_{\rm III})^{2}}{2\sigma_{\rm III}^{2}}}$ & $A_{\rm III}$ & \phantom{-}0 & \phantom{0}10 & -50& 200 \\ 
 & & $B_{\rm III}$      & \phantom{-}0 & \phantom{00}2 &-50 & 200 \\
 & & $D_{\rm III}$      & \phantom{-}0 & 200         &0 & 200 \\
 & & $\mu_{\rm III}$    & \phantom{-}2 & \phantom{0}18 &0 & 50 \\ 
 & & $\sigma_{\rm III}$ & \phantom{-}0.2 & \phantom{00}4  &0 & 20 \\
\hline
IV. & $A_{\rm IV} + B_{\rm IV} \, x_{i} + C_{\rm IV} \, x_{i}^{2} + \frac{D_{\rm IV}}{\sigma_{\rm IV} \, \sqrt{2 \pi}} e^{- \frac{(x_{i}-\mu_{\rm IV})^{2}}{2\sigma_{\rm IV}^{2}}}$ & $A_{\rm IV}$ & \phantom{-}0 & \phantom{0}10 & -50&200 \\ 
 & & $B_{\rm IV}$      & \phantom{-}0 & \phantom{00}2  &-50 & 200 \\
 & & $C_{\rm IV}$      & \phantom{-}0 & \phantom{00}0.5  &-50 & 200 \\
 & & $D_{\rm IV}$      & \phantom{-}0 & 200         &0 & 200 \\
 & & $\mu_{\rm IV}$    & \phantom{-}2 & \phantom{0}18 &0 & 50\\ 
 & & $\sigma_{\rm IV}$ & \phantom{-}0.2 & \phantom{00}4&0 & 20\\
\hline
\end{tabular}
\end{center}
\caption{Summary of the models fitted to the data, along with the
  ranges allowed for the parameters. }
\label{tab:models}
\end{table}%

The models were fitted one at a time. Two different fitting approaches were used:
\begin{enumerate}
\item[1)]  The fitting was performed using the gradient-based fitter MIGRAD from the MINUIT package~\cite{ref:Minuit} accessed from within the Bayesian Analysis Toolkit (BAT)~\cite{ref:BAT}.  The starting values for the parameters were chosen at the center of the allowed ranges given in Table~\ref{tab:models}.
\item[2)] The Markov Chain Monte Carlo (MCMC) implemented in BAT was run with its default settings first, and then MIGRAD was used to find the mode of the posterior using the parameters at the mode found by the MCMC as starting point.
\end{enumerate}

Since it is known that the distributions of the data are Gaussian and flat priors were used for the parameters, the maximum of the posterior probability corresponds to the minimum of  $\chi^2$. The $p$-values were therefore extracted for each model fit using the $\chi^2$ probability distribution as discussed in Sec.~\ref{sec:chi2gauss}.
\begin{figure}[h!tbp] 
   \centering
   \includegraphics[width=5.4in]{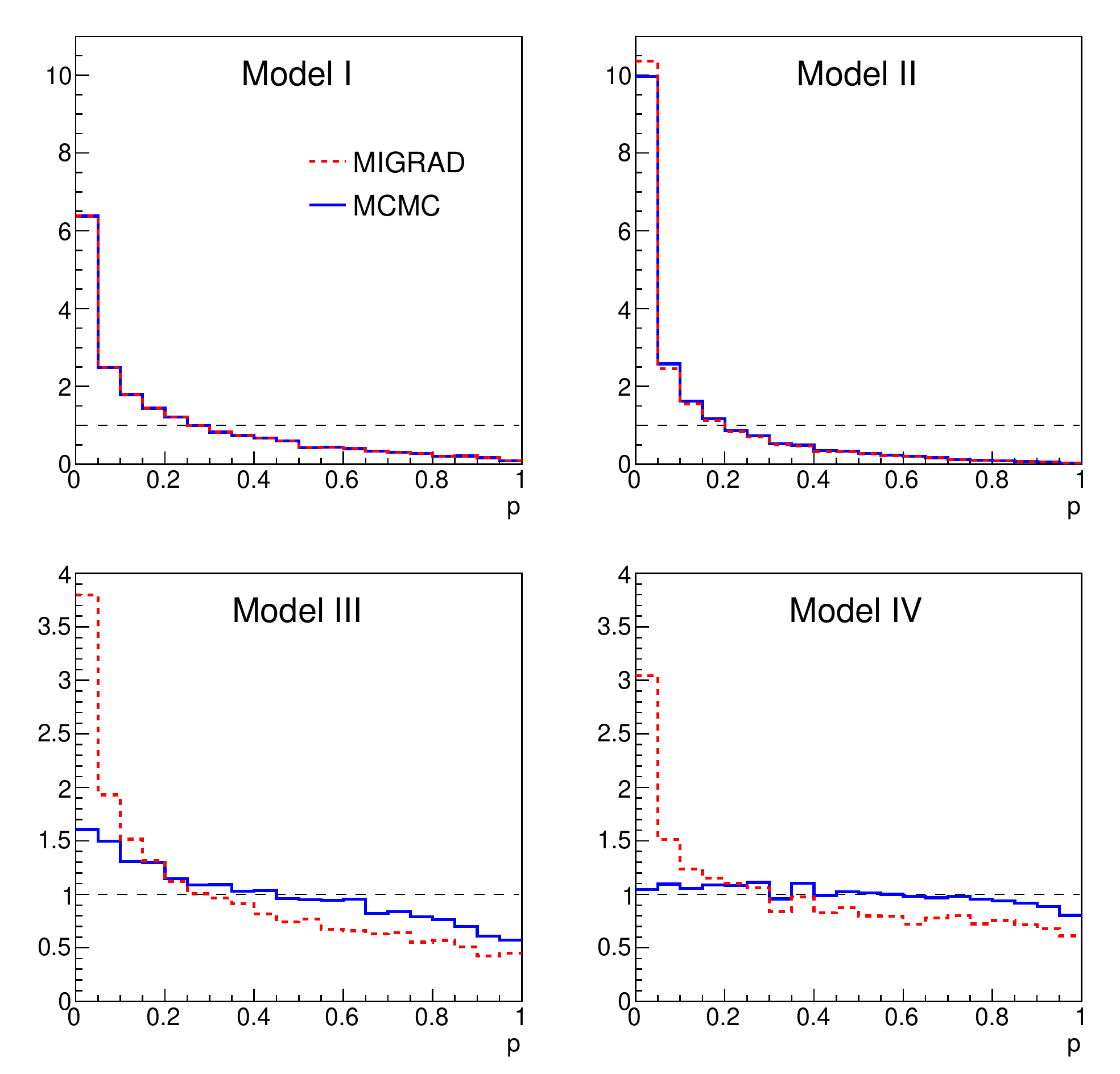}
   \caption{$p$-value distributions based on $R_{G}$. The parameters
     of the models discussed in Table~\ref{tab:models} are fitted to
     $N=25$ data points and allowing parameter
     values in the {\it small} range.  The $p$-values correspond to
     $N-n$ degrees of freedom, where $n$ is the number of fitted
     parameters.}
   \label{fig:gausschi2-25}
\end{figure} 

\subsubsection{Comparison of $p$-values}
\label{sec:comparisonpvalues}
\begin{figure}[h!tbp] 
   \centering
   \includegraphics[width=5.4in]{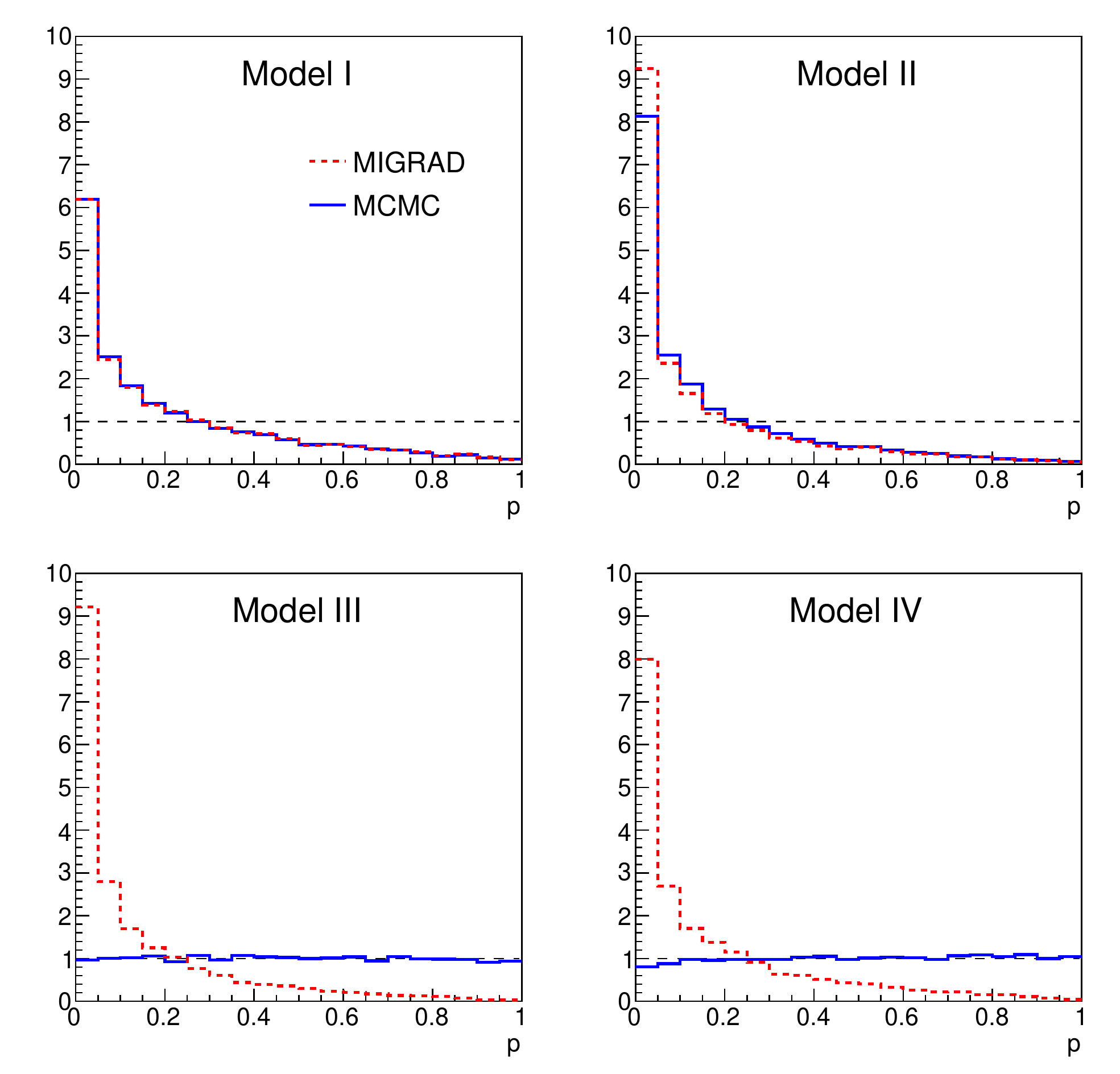}
   \caption{$p$-value distributions based on $R_{G}$. The parameters
     of the models discussed in Table~\ref{tab:models} are fitted to
     $N=25$ data points and allowing parameter values in the {\it
       large} range.  The $p$-values correspond to $N-n$ degrees of
     freedom, where $n$ is the number of fitted parameters.}
   \label{fig:gausschi2-25-large}
\end{figure}

The $p$-value distributions for the four models are given in Fig.~\ref{fig:gausschi2-25} for the case $N=25$ and using the small fit range from Table~\ref{tab:models}.   Two different histograms are shown for each model, corresponding to the two fitting techniques described above. The distributions for models I and II are peaked at small values of $p$ for MIGRAD only and for MCMC+MIGRAD, and the models would be disfavored in most experiments. For models III and IV, there is a significant difference in the $p$-value distribution found from fitting using only MIGRAD, or MCMC+MIGRAD, with the $p$-value distribution in the latter case being much flatter.  An investigation into the source of this behavior showed that the $p$-value distribution from MIGRAD is very sensitive to the range allowed for the parameters.  This can be seen in Fig.~\ref{fig:gausschi2-25-large} where the same fits were performed but now using the large ranges for the parameters (see table).  In the case where larger parameter ranges are allowed, MIGRAD will converge to parameter values which are not at the global mode more often, and the choice of the starting point and the (initial) step size for the fit are crucial. Even in this rather simple fitting problem, it is seen that the use of the MCMC can make a significant difference in the fitting results.
  
\begin{figure}[h!tbp] 
   \centering
   \includegraphics[width=2.7in]{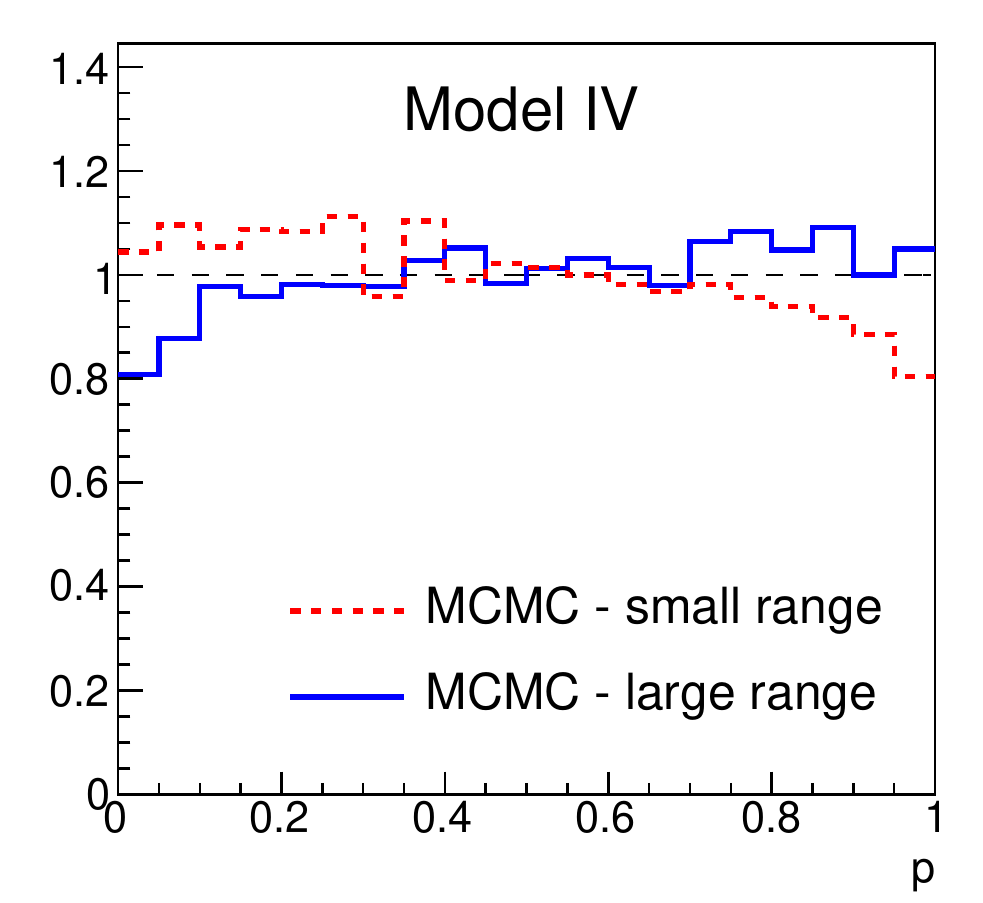}
   \caption{$p$-value distributions based on $R_{G}$ for model~IV. The
     parameters of the model are fitted to $N=25$ data points and
     allowing parameter values in the two ranges range.  The
     $p$-values correspond to $N-n$ degrees of freedom, where $n$ is
     the number of fitted parameters.}
   \label{fig:MCMCrange}
\end{figure}

 The results from the MCMC+MIGRAD fits also depended on the fit range, although to a lesser extent.   Figure~\ref{fig:MCMCrange} shows the $p$-value distributions from $\chi^2$ fits to $N=25$ data points for Model IV for the two parameter ranges.  There are small but nevertheless significant differences in the distribution, indicating that the parameter range is also an important factor in the $p$-value distribution.  Note that we are not expecting flat $p$-value distributions since the correction for the number of fitted parameters is only valid for models linearly dependent on the parameters.  This is not the case here since we have a Gaussian term in the model (see \ref{eq:true function}).

   Focusing on the results from the MCMC+MIGRAD fits, models III and IV give flat $p$-value distributions and would have a large DoB in a majority of experiments. In general, these distributions satisfy our expectations and these $p$-values can be used to update our DoBs in the models as described in Section~\ref{sec:pBayes}. 

\begin{figure}[h!tbp] 
   \centering
   \includegraphics[width=5.4in]{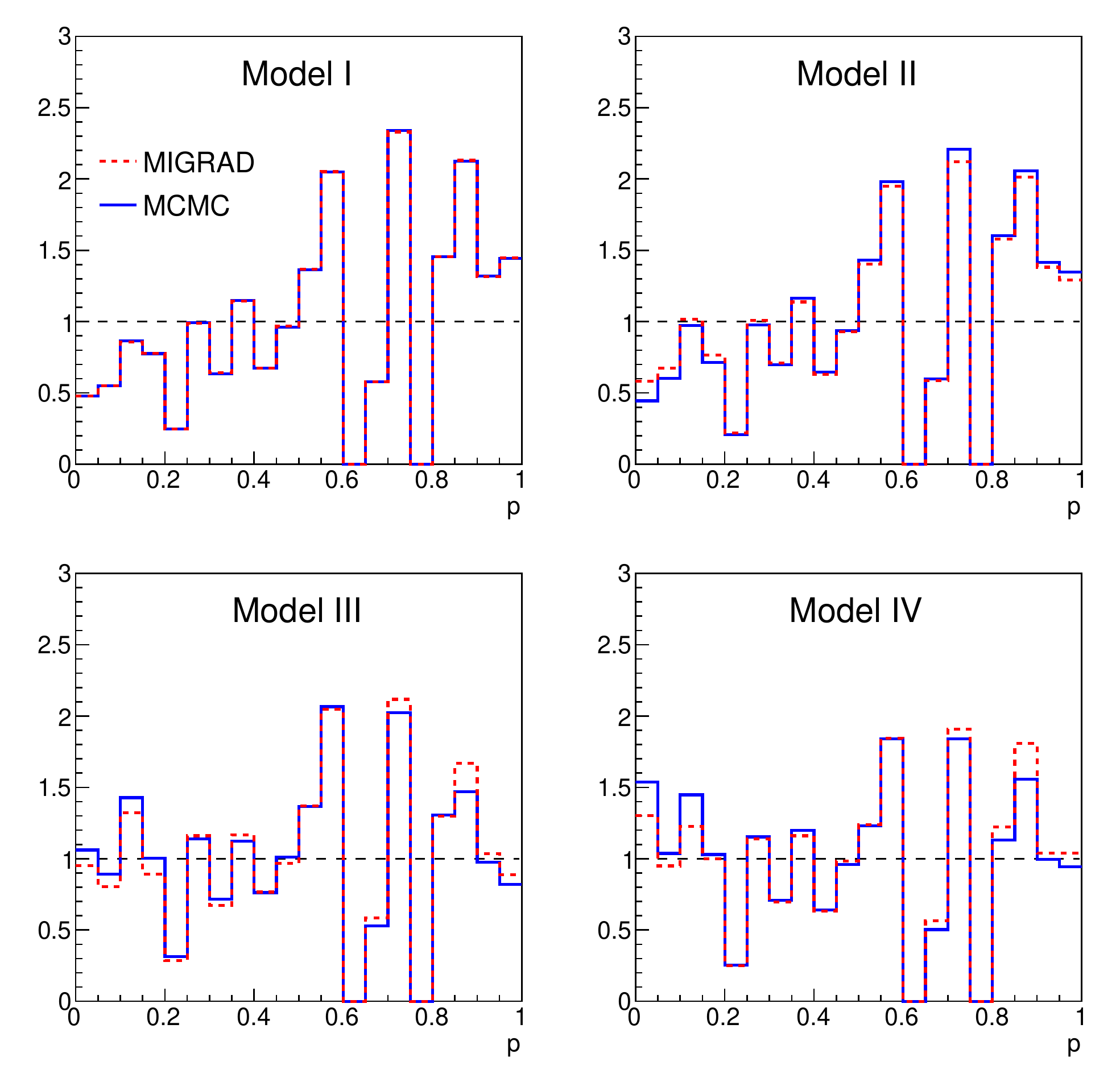}
   \caption{$p$-value distributions based on $R_{J}$. The parameters
     of the models discussed in Table~\ref{tab:models} are fitted to
     $N=25$ data points and allowing parameter values in the {\it
       small} range.  The $p$-values correspond to $N_{b}-1$ degrees
     of freedom, where $N_{b}=5$ is the number of bins.}
   \label{fig:gaussjohnson-25}
\end{figure}

The $p$-value distributions for the four models and using the Johnson discrepancy variable are given in Fig.~\ref{fig:gaussjohnson-25}.  Note that in the results shown for this descrepancy variable, we have used $\vec{\theta}^*$ rather than sampling $\vec{\theta}$ according to the posterior. We verified that this did not significantly change the $p$-value distribution of this discrepancy variable. The distributions show spiky behavior, which becomes much smoother as the number of data points increases.  However, there is still very little discriminating power between the models using this discrepancy variable.  This was generally the case for other examples considered, and we therefore do not show any further results from the Johnson discrepancy variable in this paper. 

For all further results presented in this paper, we have performed MIGRAD only fits but with starting parameter values located near the global mode (whose location can be estimated since the underlying model is known).  Only the smaller fit ranges were used. In this way, we approximate the results which would be achieved with optimal fitting.  This allows us to generate and analyze larger data sets, and thus have more sensitivity to the shape of the $p$-value distributions.

\begin{figure}[h!tbp] 
   \centering
   \includegraphics[width=5.4in]{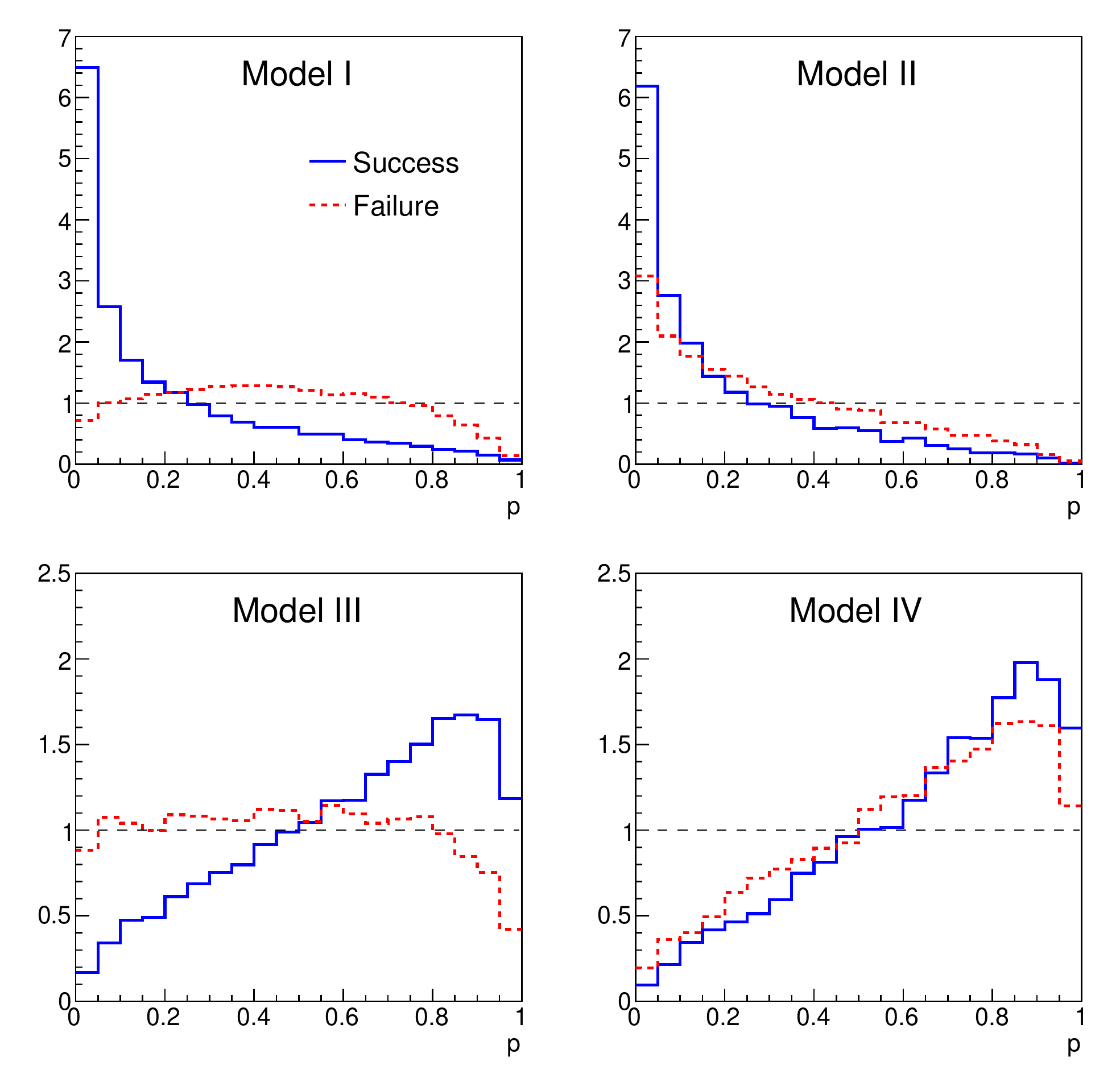}
   \caption{$p$-value distributions based on $R_{sr}$ and
     $R_{fr}$. The parameters of the models discussed in
     Table~\ref{tab:models} are fitted to $N=25$ data points and
     allowing parameter values in the {\it small} range.}
   \label{fig:runs-25}
\end{figure}

\begin{figure}[h!tbp] 
  \centering
  \includegraphics[width=3.5in]{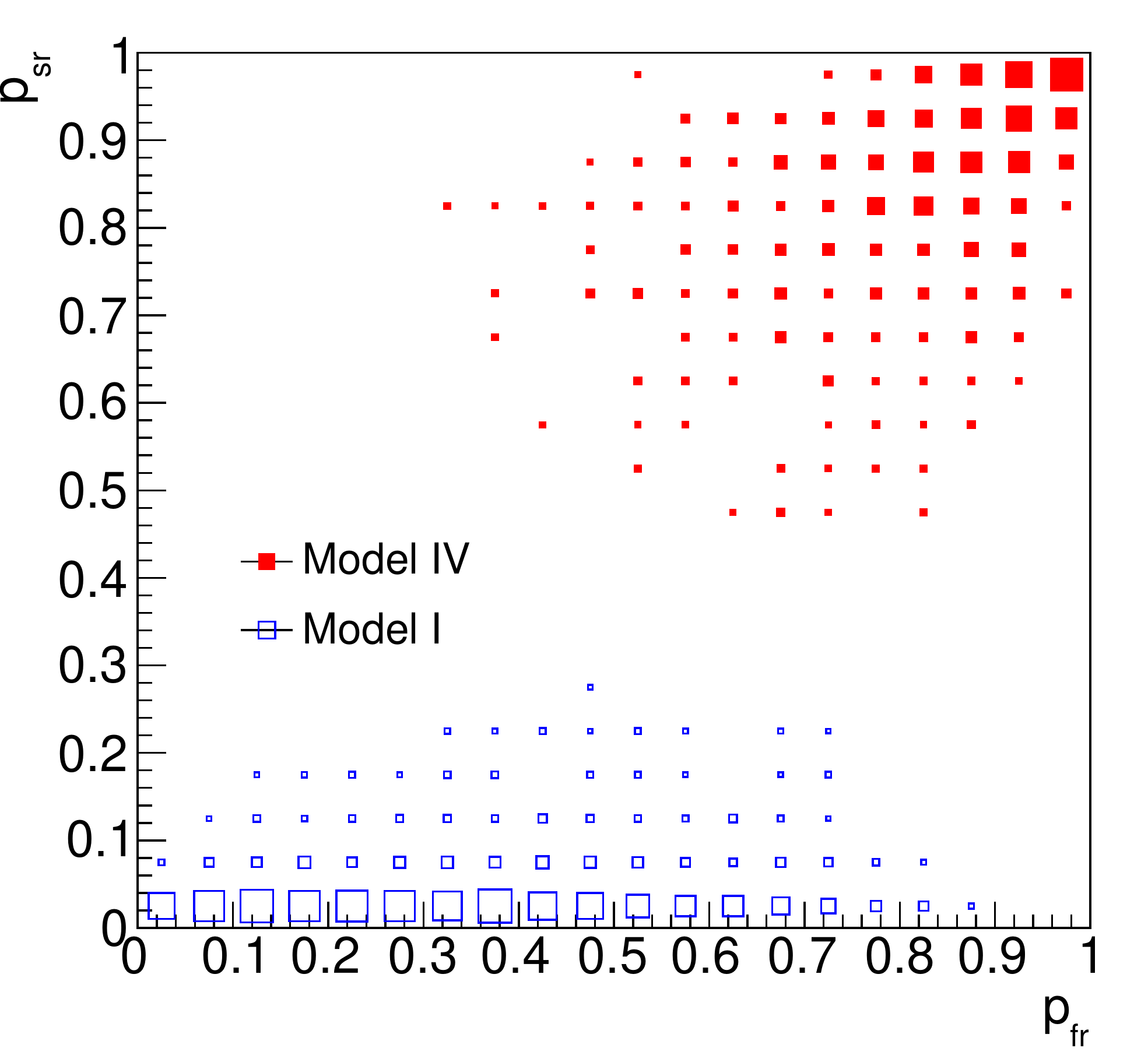}
  \caption{Joint distribution of the $p$-values for success and
    failure runs. The results are for $N=25$ data points.  Marker
    sizes in each bin are chosen relative to the bin with the highest
    probability of the individual model. Bins with probability less
    than $3.5 \cdot 10^{-3}$ have been excluded from the plot for the
    purpose of clarity.  }
  \label{fig:runs-scatter}
\end{figure}

The $p$-value distributions for the four models and using the `runs'  discrepancy variable are given in Fig.~\ref{fig:runs-25}.  In this figure, there are two $p$-value distributions in each plot since we consider both cases where we have runs of data points below the expectations, and runs of data points above the expectations.  For the correct model, the joint distribution of $p(R_{sr})$ and $p(R_{fr})$ should be symmetric. The $p$-values should generally not be too small. The distribution of $p(R_{sr})$ vs. $p(R_{fr})$ is shown in Fig.~\ref{fig:runs-scatter} for the example under consideration. 
The biasing of the $p$-values resulting from the fitting of parameters is apparent.  In using $R_{fr}$ and $R_{sr}$, rather large $p$-values should therefore be expected for valid models.

\begin{figure}[h!tbp] 
   \centering
   \includegraphics[width=5.4in]{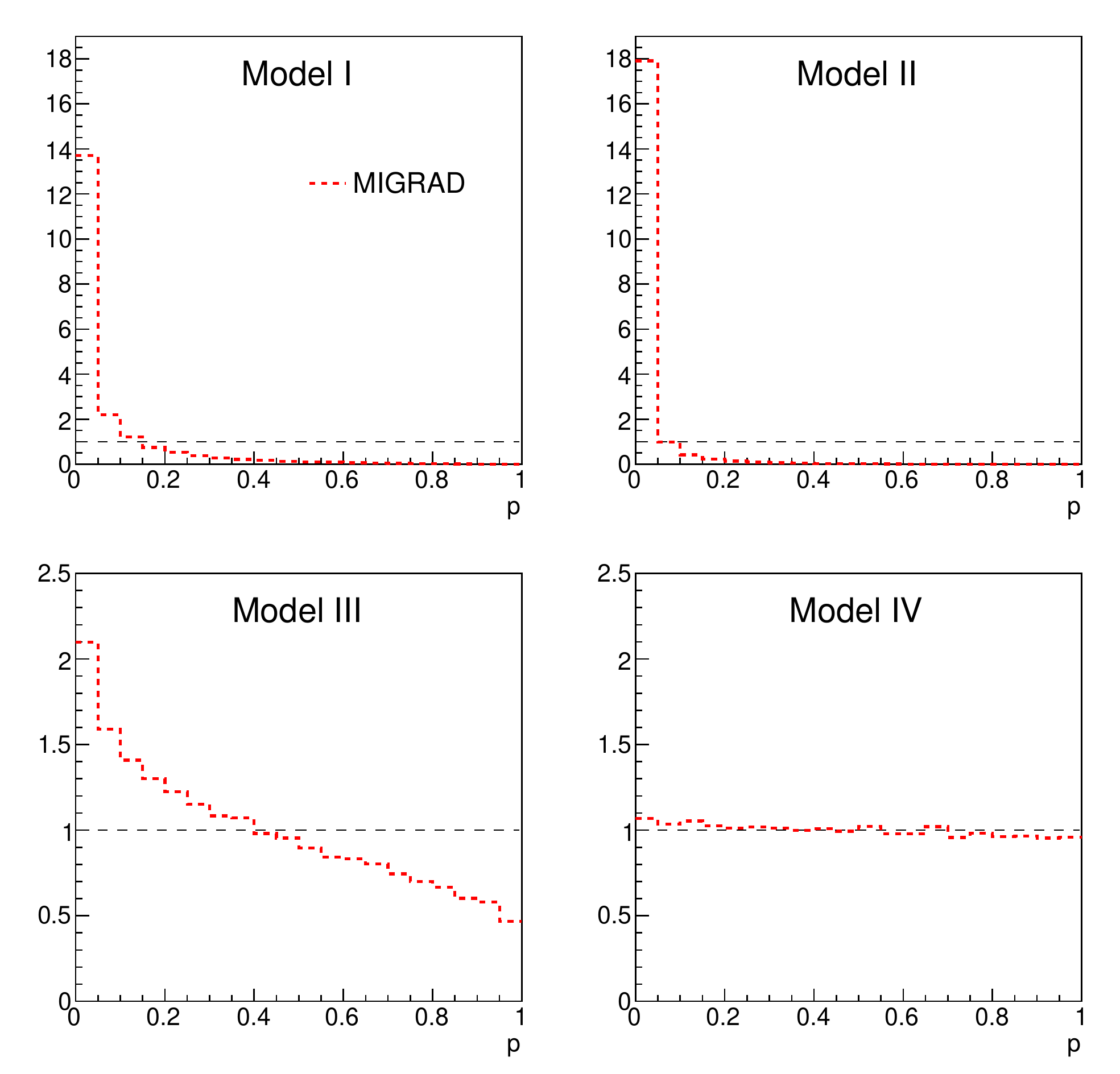}
   \caption{$p$-value distributions based on $R_{G}$. The parameters
     of the models discussed in Table~\ref{tab:models} are fitted to
     $N=100$ data points and allowing parameter
     values in the {\it small} range.  The $p$-values correspond to
     $N-n$ degrees of freedom, where $n$ is the number of fitted
     parameters.}     
   \label{fig:chi2-100}
\end{figure}

\begin{figure}[h!tbp] 
   \centering
   \includegraphics[width=5.4in]{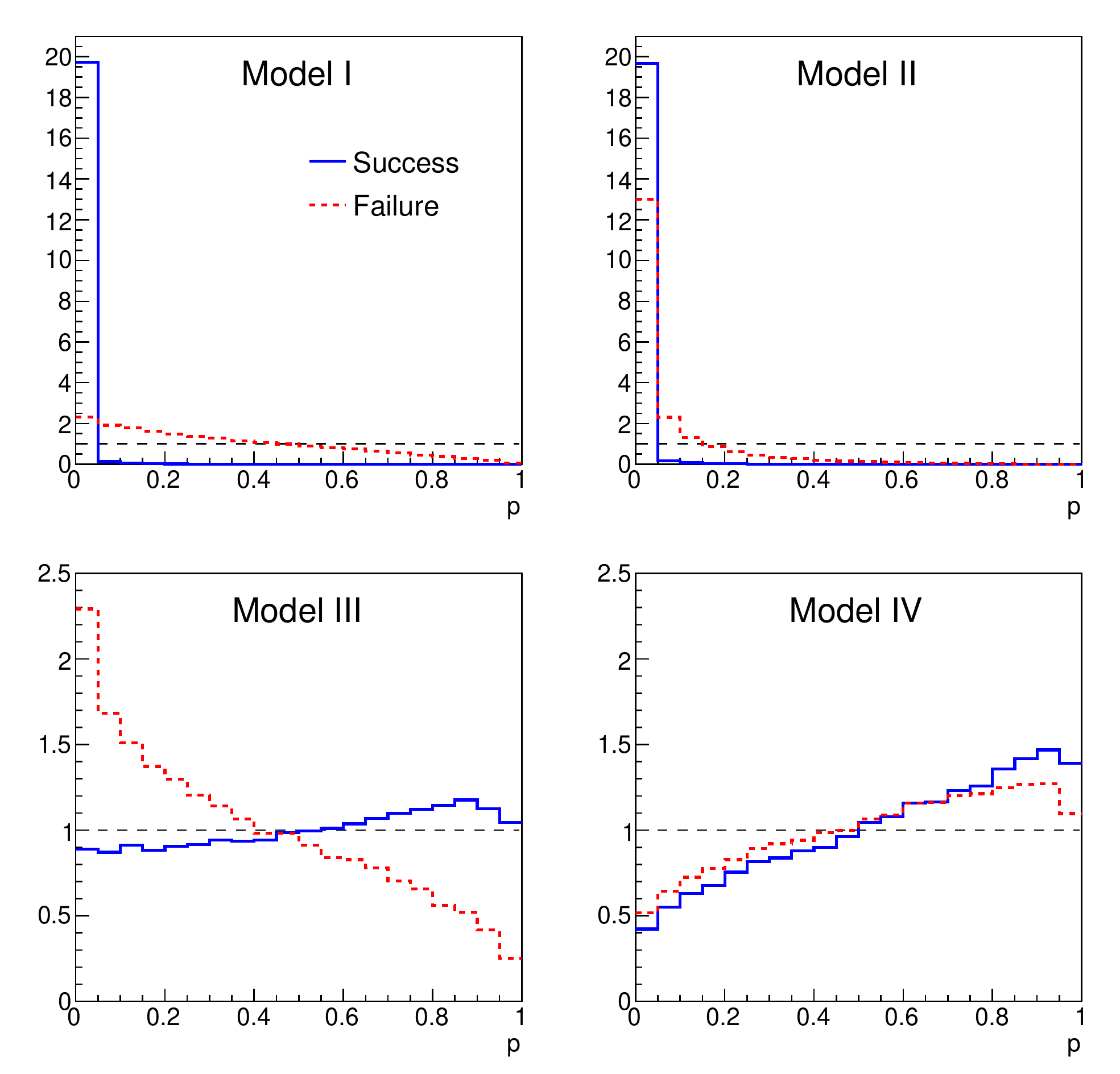}
   \caption{$p$-value distributions based on $R_{sr}$ and
     $R_{fr}$. The parameters of the models discussed in
     Table~\ref{tab:models} are fitted to $N=100$ data points and
     allowing parameter values in the {\it small} range.}
   \label{fig:runs-100}
\end{figure}

The results for the ensembles with $N=100$ data points are shown in Figs.~\ref{fig:chi2-100} and \ref{fig:runs-100}.  The models I and II now usually result in very small $p$-values using both the standard $\chi^2$ and runs discrepancy variables.  

For the $\chi^2$ fits, the $p$-value distribution for model  IV is again rather flat, as expected, but it would generally be difficult to conclude that model III is not adequate.  Although this $p$-value distribution is falling, as opposed to the case where we only had $N=25$ data points, there is still a large probability to get a sizable $p$-value. 

The runs statistic has a sharply peaked distribution near $p=0$ (for success runs) for models I,II, and should therefore have better discriminating power than the standard $\chi^2$ test.  In other words, it should more often lead to the desired result that the incorrect models are discarded.  For model III, the success runs variable behaves similarly to the $\chi^2$ test, but the failure runs variable is rather flat.  For the correct underlying model, model IV, we see that both success and failure runs variables have similar $p$-value distributions.


\newpage
\subsection{Example with Poisson uncertainties}

\subsubsection{Definition of the data}

We again use the function $f(x)$ (see \eqref{eq:true function}), but now generate data sets with fluctuations from a Poisson distribution.  The function is used to give the expected number of events in a bin covering an $x$ range, and the observed number of events follows a Poisson distribution with this mean.
For $N_{b}=25$ bins, e.g., the third bin is centered at $x_{3}=2.0$ and extends over $[1.6,\,2.4]$.
The expected number of events in this bin is defined as $\lambda_{3}=\int_{1.6}^{2.4} \frac{N_{b}}{20} f(x)\mbox{d}x$ .  

\subsubsection{Comparison of $p$-value distributions}

For the Poisson case we consider the four discrepancy variables described in sections~\ref{sec:chi2poisson}-\ref{sec:probpoiss}  to compare the $p$-value distributions for the four models (I-IV).  We also show the $p$-value distribution for the `true' model for comparison, where the $p$-value is calculated by evaluating $R=R(\vec{x}|\vec{\theta}_{\rm true},M)$. 

\begin{figure}[h!tbp] 
   \centering
   \includegraphics[width=3.5in]{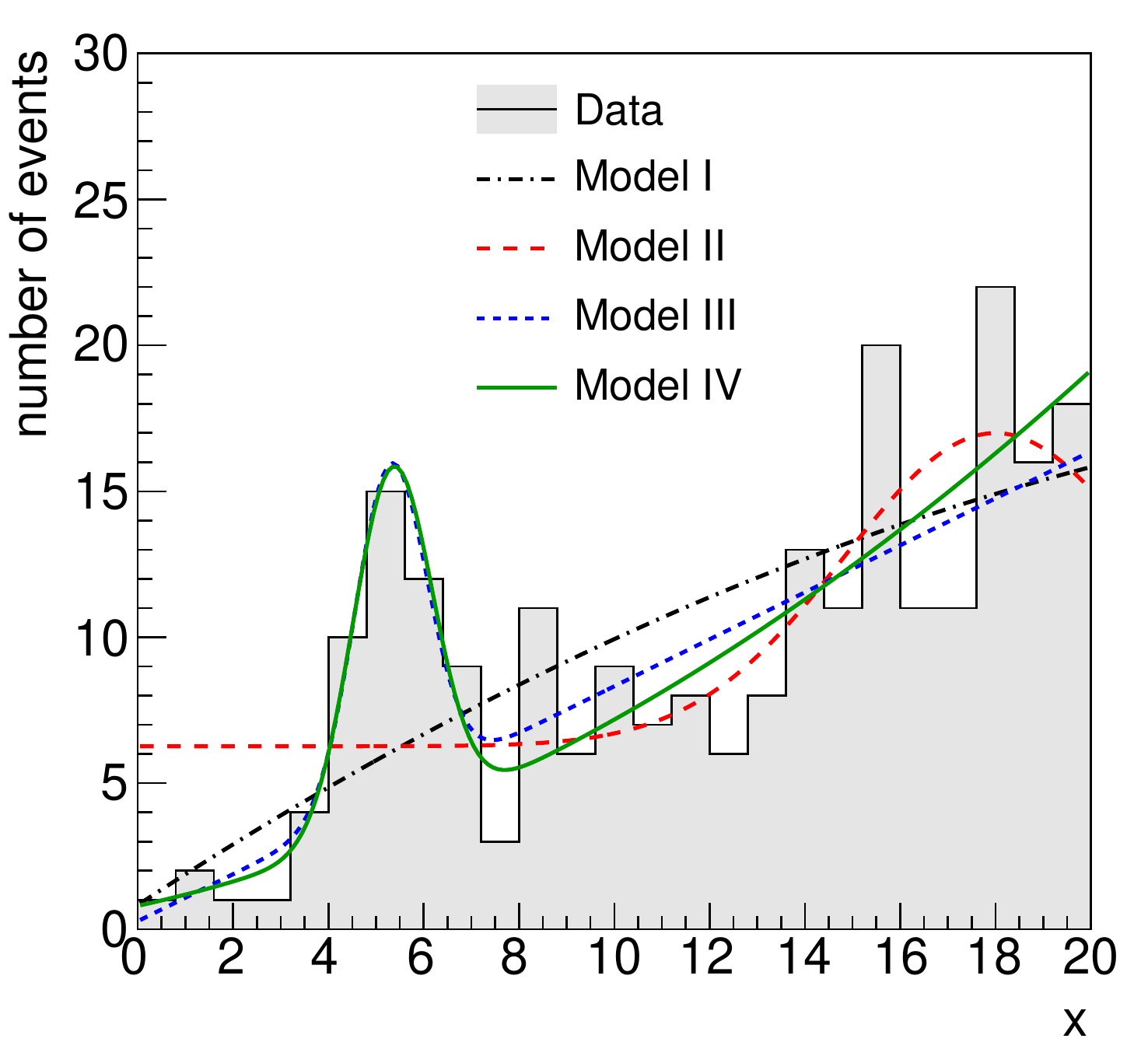}
   \caption{Example fits to one data set for the four models defined in Table~\ref{tab:models} for $N_{b}=25$ and the {\it small} parameter range.  The lines shows the fit functions evaluated using the best-fit parameter values.}
   \label{fig:fitpoiss}
\end{figure}

\begin{figure}[h!tbp] 
   \centering
   \includegraphics[width=5.4in]{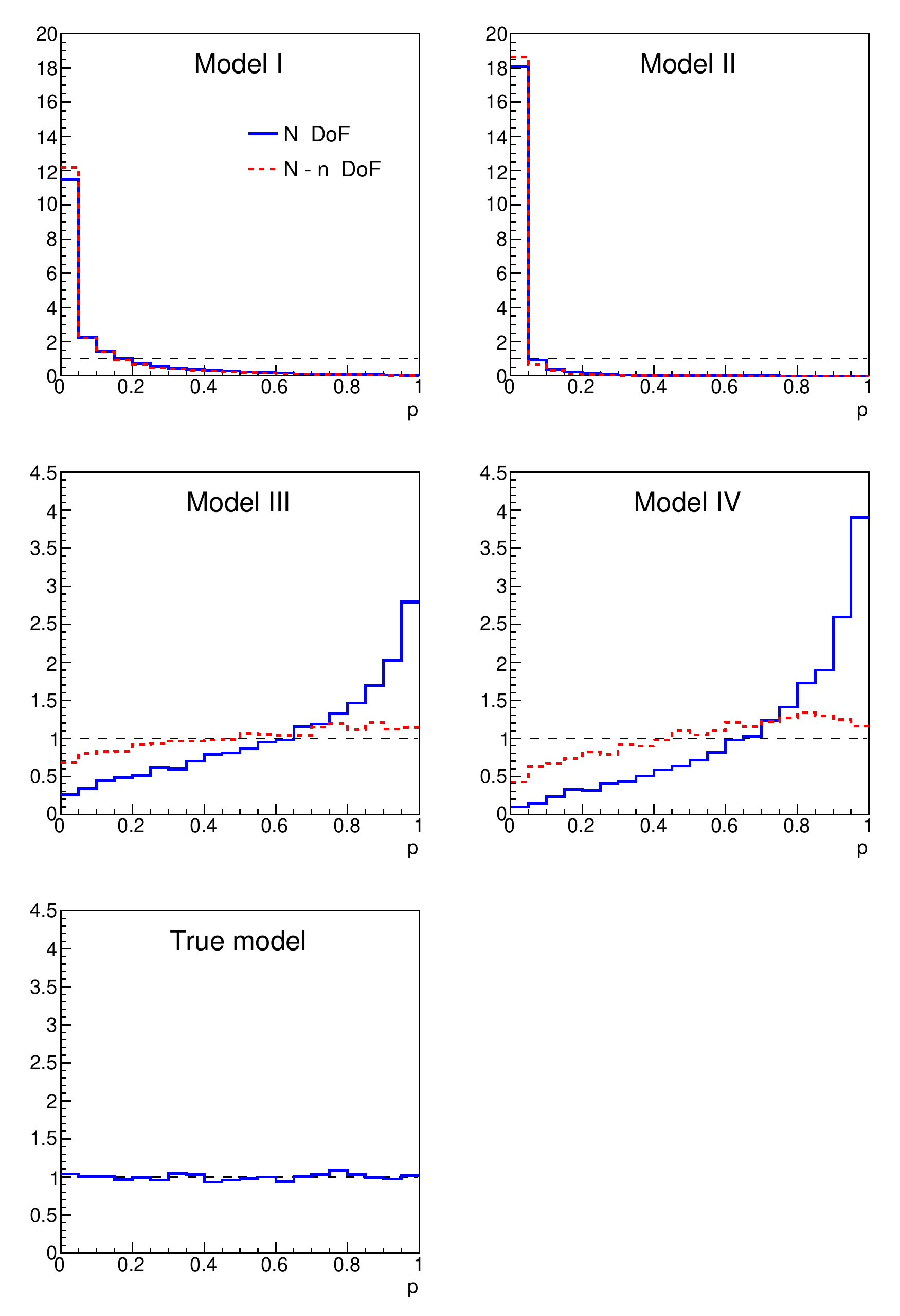}
   \caption{$p$-value distributions based on $R_{L}$. The parameters
     of the models discussed in Table~\ref{tab:models} are fitted to
     $N_{b}=25$ bins and allowing parameter values in the {\it small}
     range.  The $p$-values correspond to $N_{b}$ and $N_{b}-n$
     degrees of freedom, where $n$ is the number of fitted parameters.}     
   \label{fig:probpoiss}
\end{figure}

Different fits were performed for each of the different discrepancy variable definitions.  For the $\chi^2$ discrepancy variables, $\chi^2$ minimization was performed using either the expected or observed number of events as weight.  For the likelihood ratio test, a maximum likelihood fit was performed.  The likelihood was defined as
\begin{equation}
P(\vec{m}|\vec{\theta}, M) = \prod_{i=1}^{N_{b}} \frac{e^{-\lambda_i}\lambda_i^{m_i}}{m_i!}
\end{equation}
where $m_i$ is the observed number of events in a bin and $\lambda_i=\lambda_i(\vec{\theta})$.  Since we use flat priors for the parameters,
the same results were used for the case where the discrepancy variable is the probability of the data. A typical likelihood fit result is shown in Fig.~\ref{fig:fitpoiss} together with the data, while the $p$-value distributions are given in Figs.~\ref{fig:chi2exp}-\ref{fig:probpoiss}.

For all definitions of $p$-values considered here, models I,II generally lead to low DoBs, while models III and IV generally have high DoBs.  However, there are differences in the behavior of the $p$-values which we now discuss in more detail.

The lower left panel in Figs.~\ref{fig:chi2exp}-\ref{fig:probpoiss}
show the $p$-value distribution taking the true model and the true
parameters, and can be used as a gauge of the biases introduced by the
$p$-value definition.  There is a bias for all discrepancy variables
shown in Fig.~\ref{fig:chi2exp} because approximations are used for
the probability distribution of the discrepancy variable (using
$P(\chi^2|N)$ rather than $P(R_{P}|N_{b})$ or $P(R_{N}|N_{b})$. This
  bias is usually small, except in the case where the Neyman $\chi^2$
  discrepancy variable is used.  In this case, the probability of a
  very small $p$-value is much too high even with the true parameters.
  The $p$-value distribution using the probability of the data, on the
  other hand, is flat when the true parameters are used as seen in
  Fig.~\ref{fig:probpoiss}, so that this choice would be optimal in
  cases where no parameters are fit.

\subsubsection{Pearson $\chi^2$ and Neyman $\chi^2$}
\begin{figure}[h!tbp] 
   \centering
   \includegraphics[width=5.4in]{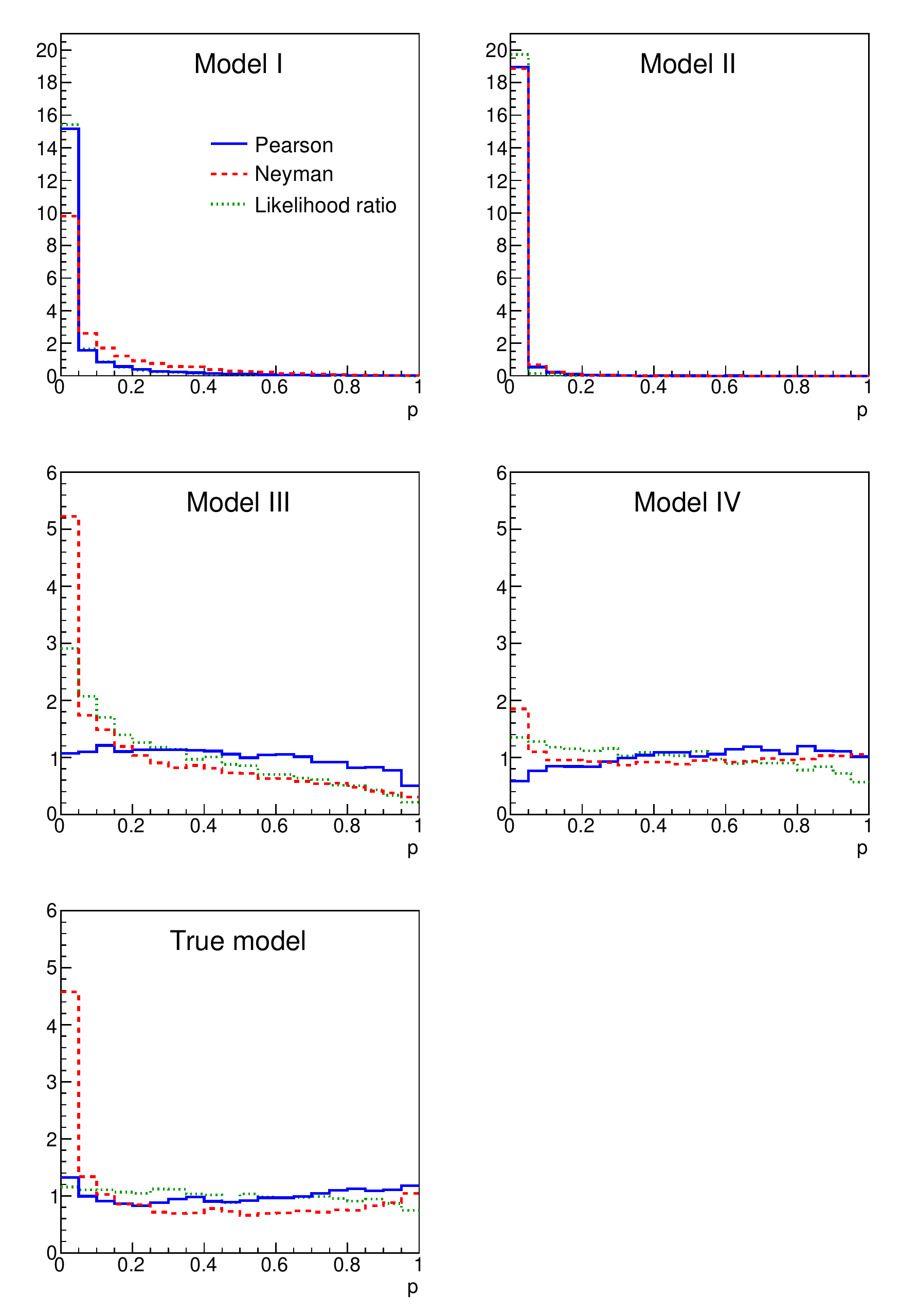}
   \caption{$p$-value distributions based on $R_{P}$, $R_{N}$ and
     $R_{C}$. The parameters of the models discussed in
     Table~\ref{tab:models} are fitted to $N_{b}=25$ bins and allowing
     parameter values in the {\it small} range.  The $p$-values
     correspond to $N-n$ degrees of freedom, where $n$ is the number
     of fitted parameters.}     
   \label{fig:chi2exp}
\end{figure}

Comparison of the behavior of the Pearson $\chi^2$ to the Neyman $\chi^2$,  Fig.~\ref{fig:chi2exp}, clearly shows that using the expected number of events as weight is superior.  The spike at 0 in the $p$-value distribution when using the observed number of events indicates that this quantity does not behave as expected for a $\chi^2$ distribution when dealing with small numbers of events, and will lead to undesirable conclusions more often than anticipated.  The behavior of the Pearson $\chi^2$ using the expected number of events for the different models is quite satisfactory, and this quantity makes a good discrepancy variable even in this example with small numbers of events in the bins. Both of these $p$-value distributions become flat in the case of large number of events in all bins.

\subsubsection{Likelihood Ratio}
While models I,II are strongly peaked at small $p$-values, for this definition of a $p$-value models III,IV also have a falling $p$-value distribution.  This is somewhat worrisome.  In assigning a DoB to models based on this $p$-value, this bias towards smaller $p$-values should be considered, otherwise good models will be assigned too low a DoB.  The behavior of the $p$-value extracted from Pearson's $\chi^2$ is preferable to the likelihood ratio, as it will lead to value judgments more in line with expectations.

\subsubsection{Probability of the Data}
Using this definition, the $p$-value distribution is flat for the true model, since this is the only case where the correct probability distribution of the discrepancy variable is employed.  For the fitted models, two $p$-value distributions are shown - one uncorrected for the number of fitted parameters, and one corrected for the fitted parameters as described in Section~\ref{sec:probpoiss}.  Using the uncorrected $p$-values, both models III and IV show $p$-value distributions peaking at $p=1$, so that both models would typically be kept.  As expected, the correction for the number of fitted parameters pushes $p$ to lower values, and produces a nearly flat distribution for model IV.  This ad-hoc correction works well for this example.  In general, this $p$-value definition has similar properties to the Pearson $\chi^2$ statistic, but with the advantage of a flat $p$-value distribution when the true model was used and no parameters were fit.

\subsection{Exponential Decay}
\label{sec:expo}

When dealing with event samples with continuous probability distributions for the measurement variables, it is common practice when determining the parameters of a model to use a product of event probabilities (unbinned likelihood):
$$P(\vec{x}|\vec{\theta},M)=\prod_{i=1}^{N}P(x_i|\vec{\theta},M) \;\; .$$
If the model chosen is correct, then this definition for the probability of the data (or likelihood) can be used successfully to determine appropriate ranges for the parameters of the model.  However, $P(\vec{x}|\vec{\theta},M)$ defined in this way has no sensitivity to the overall shape of the distribution and can lead to unexpected results if this quantity is used in a GoF test of the model in question.  We use a common example, exponential decay, to illustrate this point (see also discussions in ~\cite{ref:Eadie} and ~\cite{ref:exp}).

Our model is that the data follows an exponential decay law.  We measure a set of event times, $\vec{t}$, and analyze these data to extract the lifetime parameter $\tau$.  We define two different probabilities of the data $\vec{D}=\{ t_i\}\; \; (i=1\dots N) $
\begin{enumerate}
\item Unbinned likelihood \[
P\left(\vec{D}|\tau\right)=\prod_{i=1}^{N}\frac{1}{\tau}e^{-t_{i}/\tau}\; , \]

\item Binned likelihood \[
P(\vec{D}|\tau)=\prod_{i=1}^{N_{b}}\frac{\lambda_{i}^{m_{i}}}{m_{i}!}e^{-\lambda_{i}},\,\,\,\,\, \lambda_{i}\left(\tau\right)=\int_{t_{i}}^{t_{i}+\Delta t}
\mbox{d}t\,\,\frac{N}{\tau}e^{-t/\tau}\; .\]

\end{enumerate}
In the first case, the probability density is a product of the densities for the individual events, while in the second case the events are counted in time intervals and Poisson probabilities are calculated in each bin.  The expected number of events is normalized to the total number of observed events.  We consider time intervals with a width $\Delta t=1$ unit and time measurements ranging from $t=0$ to $t=20$.  The overall probability is the product of the bin probabilities.  For each of these two cases, we consider the $p$-value determined from the distribution of $R=P(\vec{D}|\tau)$.

\begin{figure}[h!tbp] 
   \centering
   \includegraphics[width=3.5in]{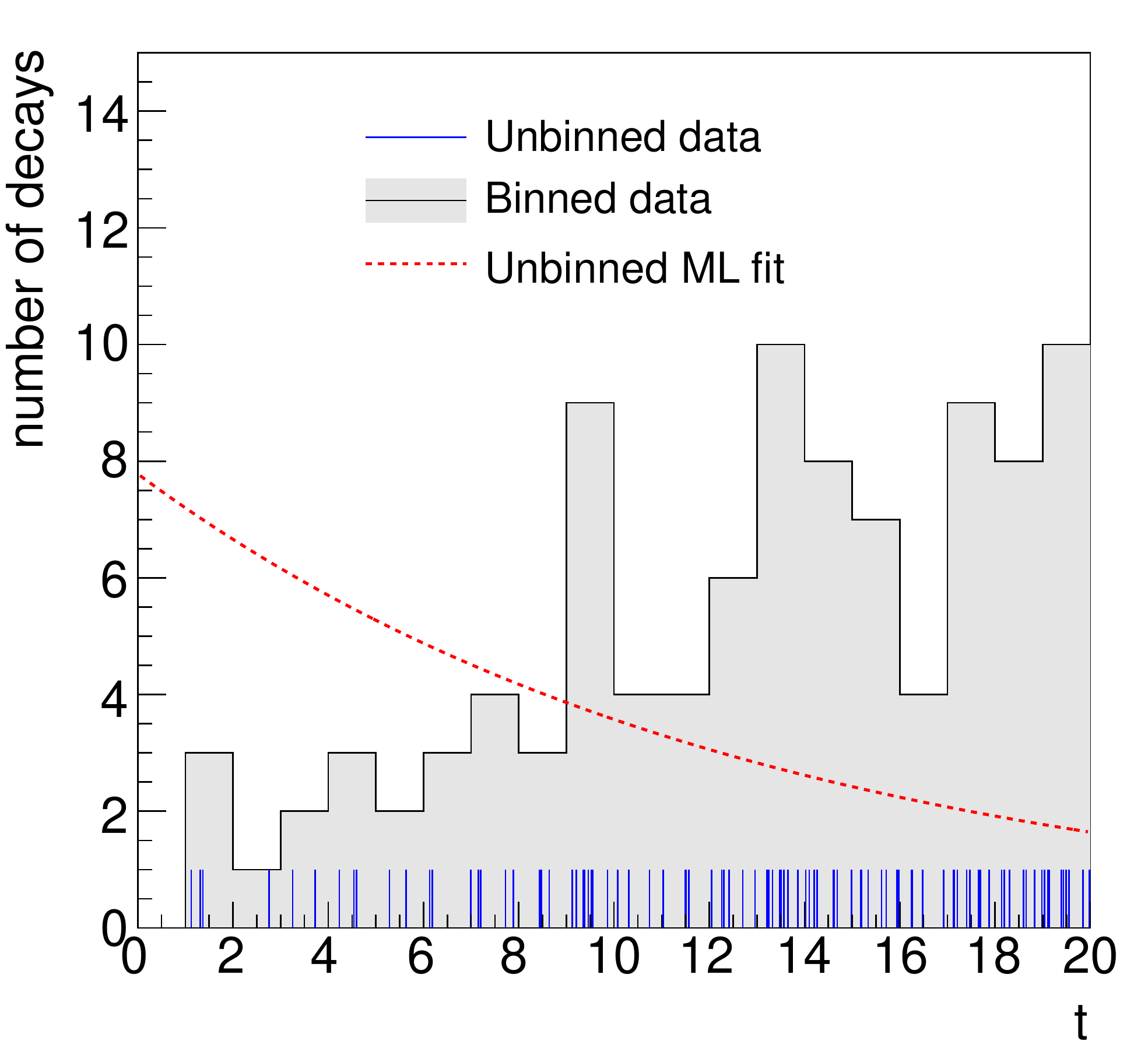}
   \caption{Typical data set used for the fitting of the exponential model.  The individual event times are shown, as well as the binned contents as defined in the text. The best fit exponential from the unbinned likelihood is also shown on the plot, normalized to the total number of fitted events.}     
   \label{fig:linear}
\end{figure}

In order to make the point about the importance of the choice for the discrepancy variable for GoF tests, we generated data which do not follow an exponential distribution.  The data is generated according to a linearly rising function, and a typical data set is shown in Fig. ~\ref{fig:linear}.

\subsubsection{Product of exponentials}

If the data are fitted using a flat prior probability for the lifetime parameter, then we can solve for the mode of the posterior probability of $\tau$ analytically, and get the well-known result
 \[\tau^*=\frac{1}{N}\sum t_{i}\; .\]

Defining $\xi \equiv\sum t_{i}$, so $\tau^*=\xi/N$, we can also solve analytically for the 
$p$-value:

\begin{eqnarray*}
p &  = & \int_{\sum t_{i}'>\xi}\mbox{d}t_{1}'\int\mbox{d}t_{2}'\dots\left(\tau^*\right)^{-N}e^{-\sum t_{i}'/\tau^*}\end{eqnarray*}

and the result is \[
p=1-P\left(N,N\right)\]
with the regularized incomplete Gamma function \[
P\left(s,x\right)=\frac{\gamma\left(s,x\right)}{\Gamma\left(s\right)}=\frac{\int_{0}^{x}t^{s-1}e^{-t}\mbox{d}t}
{\int_{0}^{\infty}t^{s-1}e^{-t}\mbox{d}t}\; . \]
Surprisingly, as $N$ increases $p$ is approximately constant with a value $p\approx 0.5$. Regardless
of the actual data, $p$ is never small and depends only on $N$. The best fit exponential is compared to the data in Fig.~\ref{fig:linear}, and yields a large $p$-value although the data is not exponentially distributed.  This $p$-value definition is clearly useless, and the unbinned likelihood is seen to not be appropriate as a GoF statistic.

The definition of $\chi^2$ in the Gaussian data uncertainties case also results from a product of probability densities, so the question as to what is different in this case needs clarification.  The point is that in the case of fitting a curve to a data set with Gaussian uncertainties, the data are in the form of pairs, $\{(x_i,y_i)\}$, where the measured value of $y$ is assigned to a particular value of $x$, and the discrepancy between $y$ and $f(x)$ is what is tested.  In other words, the value of the function is tested at different $x$ so the shape is important.  In the case considered here, the data are in the form $\{x_i\}$, and there is no measurement of the value of the function $f(x)$ at a particular $x$.  The orderings of the $x_i$ is irrelevant, and there is no sensitivity to the shape of the function.

\subsubsection{Product of Poisson probabilities}

In this case, the $p$-value for the model is determined from the product of Poisson probabilities using event counting in bins as described above.  Since the expected number of events now includes an integration of the probability density and cover a wide range of expectations, it is sensitive to the distribution of the events and gives a valuable $p$-value for GoF.  The $p$-value using $R_{P}$ for the data shown in Fig.~\ref{fig:linear} is $p=0$ within the precision of the calculation.  The exponential model assumption is now easily ruled out.  

In comparison to the unbinned likelihood case, the data are now in the form $\vec{m}$, where the $m_i$ now refer to the number of events in a well-defined bin $i$ which defines an $x$ range.  In other words, we have now a measurement of the height of the function $f(x)$ at particular values of $x$, and are therefore sensitive to the shape of the function.

As should be clear from this example, the choice of discrepancy
variable can be very important in GoF decisions: Maximum Likelihood
estimation with unbinned data will always give an optimal parameter
estimate in terms of bias and variance, but it will give no
information about the correctness of the model.
\section{Discussion}

We have investigated a number of possible discrepancy variables and $p$-value definitions for Goodness-of-Fit tests, with the understanding that these $p$-values are to be used to make judgments on the acceptability of models.  The sense in which the $p$-values are used follows Bayesian logic as described in Section~\ref{sec:pBayes}. For this purpose, the $p$-value distributions should be reasonably uniform for models which are considered good and they should peak at small values for poor models.  Using these requirements to guide us, we classify our results in Table~\ref{tab:results} for the example data sets and models described in the text.  Based on our experience with these examples, we also summarize our suggestions for the use of the different discrepancy variables and $p$-value definitions considered here.

\begin{sidewaystable}[htdp]
\begin{center}
\begin{tabular}{|c|l|l|l|c|}
\hline
Discrepancy variable & $p$-value definition & Data type & Comment &    \\
\hline
$R_{G}$    & $P(\chi^{2}>R_{G}^{D}|N-n)$ & $\vec{D}=\{(x_i,y_i)\}$                   & Flat if $n=0$ or model linear & $+$ \\
\hline
$R_{sr/fr}$ & $P(R_{sr/fr}>R_{sr/fr}^{D}|N)$ & $\vec{D}=\{(x_i,y_i)\}$             & Generally not flat for $n>0$ & $+$ \\
\hline 
$R_{P}$    & $P(\chi^{2}>R_{P}^{D}|N_{b}-n)$ & $\vec{D}=\vec{m}$                     & Reasonably flat & $+$ \\
$R_{N}$    & $P(\chi^{2}>R_{N}^{D}|N_{b}-n)$ & $\vec{D}=\vec{m}$                     & Peaked at 0 for small numbers of events & $-$ \\
\hline
$R_{C}$    & $P(\chi^{2}>R_{C}^{D}|N_{b}-n)$ & $\vec{D}=\vec{m}$                     & Distribution for $R_{C}$ is less flat than that for $R_{P}$ & $+$ \\
\hline
$R_{L}$    & $P(R_{L}<R_{L}^{D}|\vec{\theta}^{*},M)$ & $\vec{D}=\{(x_i,y_i)\}$    & Same as $R_{G}$ & $+$ \\
$R_{L}$    & $P(R_{L}<R_{L}^{D}|\vec{\theta}^{*},M)$ & $\vec{D}=\vec{m}$          & Almost flat with correction & $+$ \\
$R_{L}$    & $P(R_{L}<R_{L}^{D}|\vec{\theta}^{*},M)$ & $\vec{D}=\{x_i\}$        & No sensitivity to shape & $-$ \\
\hline
$R_{J}$    & $P(\chi^{2}>R_{J}^{D}|N_{b}-1)$ & $\vec{D}=\{(x_i,y_i)\}$               & Spiky distribution & $-$ \\
$R_{J}$    & $P(\chi^{2}>R_{J}^{D}|N_{b}-1)$ & $\vec{D}=\vec{m}$                     & Spiky distribution & $-$ \\
\hline
\end{tabular}
\end{center}
\caption{Summary of the discrepancy variables considered in this paper
  and their performance for different data types. Here, $n$ is the
  number of fit parameters, $N$ is the number of events and $N_{b}$
  denotes the number of bins. The comment concerns the shape of the
  $p$-value distribution for good models.  The last column gives our
  recommendation for use as a discrepancy variable for model
  judgments based on the given $p$-value definition.}
\label{tab:results}
\end{sidewaystable}%

Most common  $p$-values use approximations for the distribution of the discrepancy variable.  This leads to non-flat distributions of the $p$-value even when no parameters are fitted, and these deviations can be severe in some cases.  We discussed the use of the probability of the data itself as a discrepancy variable, and showed that it is flat in the case of  no fitted parameters and Poisson distributed data.  This was due to the use of the correct probability distribution for the discrepancy variable.  The algorithm described in the Appendix can be used to generate the correct $p$-value distribution for any discrepancy variable for Poisson distributed data.

For composite models, where parameters of the model are fit to the data before a $p$-value is calculated, we find that $p$-value distributions can depend strongly on the technical approach used to fit the data. Using gradient-based approaches to find minima or maxima requires considerable attention from the user and generally fine tuning of fit ranges and starting values.  The fine-tuning will certainly affect the $p$-value distributions, making their use more difficult. Setting range limits effectively means defining a prior, and impacts $p$-value distributions. First fitting the data with a Markov Chain Monte Carlo before using MIGRAD gave considerably better results, in the sense that the $p$-value distributions extracted in this way followed expectations. We therefore recommend using a MCMC to map out the parameter space as a start to the fitting procedure in situations where giving good starting values for a gradient-based fit is difficult.

\newpage
\begin{appendix}
\section{Fast $p$-value evaluation in MCMC for Poisson distributed data}

The peak probability for a Poisson distribution
$$P(m|\lambda)=\frac{e^{-\lambda}\lambda^m}{m!}$$
occurs for $m=\left\lfloor \lambda \right\rfloor$.  If the probability distribution of a data set is modeled as a product of Poisson terms, then the 
highest probability is given for $\{m_i\}=\{\left\lfloor \lambda_i \right\rfloor\}$. 
We use this to define the starting point for a Markov Chain, and move the bin contents up or down (chosen randomly) 
at each iteration.  For each attempted change in the bin content, we apply the usual Metropolis test~\cite{ref:Metropolis}.  The probability $P(\vec{m}|\vec{\lambda})$
is easily updated at each change.  E.g., if the result in bin $i$ increases from $m_i$ to $m'_i$, then the probability 
changes by
$$P\rightarrow P\frac{m_i!}{m'_i!}\lambda_{i}^{m'_i-m_i} \;\;.$$
A large number of experiments can be quickly simulated and the $p$-value extracted.

\end{appendix}



\begin{thebibliography}{9}
\bibitem{ref:Pearson} K. Pearson, `On the Criterion that a given System of Deviations from the Probable in the Case of a Correlated System of Variables is such that it can be reasonably supposed to have arised from Random Sampling', Philosophical Magazine, {\bf 50} (1900) 157.
\end{thebibliography}
\end{document}